\documentclass[10pt]{interact}
\usepackage{amsfonts}
\usepackage{amssymb}
\usepackage{amsmath}
\usepackage{epstopdf}
\usepackage[caption=false]{subfig}
\usepackage[numbers,sort&compress]{natbib}
\usepackage{xcolor}
\usepackage{fontsize}
\usepackage{xurl}
\usepackage[colorlinks=true,citecolor=blue,urlcolor=blue,linkcolor=blue]{hyperref}
\usepackage{braket}

\renewcommand{\textsl}[1]{{#1}}

\bibpunct[, ]{[}{]}{,}{n}{,}{,}

\theoremstyle{plain}
\newtheorem{theorem}{Theorem}[section]

\newtheorem{proposition}[theorem]{Proposition}
\theoremstyle{definition}

\theoremstyle{remark}

\input{tcilatex}
\begin{document}

\title{Decoherence, Perturbations and Symmetry in Lindblad Dynamics ---
Implications for Diffractive Dissociation}
\author{ 
\name{A.~Y. Klimenko\textsuperscript{a}\thanks{Contact A.Y.
Klimenko. Email: a.klimenko@uq.edu.au}} 
\affil{\textsuperscript{a} CMES,
SoMME, The University of Queensland, Australia \\ ------ \\
{\bf published:}  \href{https://link.springer.com/article/10.1140/epjc/s10052-026-15689-x}{EPJC, {\bf 86}, 482, (2026)}
} }

\maketitle

\begin{abstract}
We extend a perturbative Dyson-type treatment and discrete-symmetry
constraints from the Schr\"{o}dinger and von Neumann equations to a
dephasing Lindblad framework. This work develops further the odd-symmetric
formulation involving dual temporal conditions from general dynamical
considerations to specific tools of quantum mechanics. Applying the
resulting scaling relations to published single- and double-diffractive data
in $pp$ and $p\bar{p}$ collisions (ISR, UA4, UA5, CDF, D0, ALICE, and E710),
we show that single-diffraction cross sections are well described by a
three-parameter fit with a relative RMS deviation of $\sim 4\%$,
substantially improving upon conventional approximations that neglect
decoherence. The extracted decoherence factor is consistently $\phi \approx
0.89$, in agreement across SD, DD, and E710-based (direct) estimates, and is
naturally interpreted as $\phi =1$ for CP-invariant dephasing but $\phi <1$
for CPT-invariant dephasing, favouring the latter.
\end{abstract}

\begin{keywords}
dissipative quantum dynamics; decoherence and recoherence; CP and CPT symmetries; diffractive dissociation in proton collisions 
\end{keywords}

\changefontsize{10pt}

\section{Introduction}

The fundamental laws governing microscopic dynamics are (to a very good
approximation) time-reversal symmetric. This motivates considering physical
and cosmological descriptions in which the Universe is constrained not only
by a boundary condition in the remote past, but also by a second boundary
condition in the remote future. A time-symmetric cosmological scenario of
this type was discussed by Gold, who argued that in a recollapsing universe
the thermodynamic arrow may reverse during the contracting phase \cite%
{Gold1962Arrow,Price1996TimesArrow}. A systematic implementation of
independent two-time boundary conditions (i.e.\ not treating one boundary as
determined by the other or by expansion or contraction dynamics of the
Universe) was advocated and analysed by Schulman in explicit
\textquotedblleft two-time\textquotedblright\ and bridge-type constructions 
\cite{Schulman1991TwoTime,Schulman1997TimesArrowsBook}.

The conceptual importance of conditioning on$\mathbb{\ }$two temporal
endpoints has been emphasised recently by Scharnhorst, Wolpert, and Rovelli
under an \textit{epistemic} interpretation of probability: they show that
conditioning expectations on information at two times can qualitatively
change the expected entropy dynamics, leading to \textquotedblleft Boltzmann
bridge\textquotedblright\ behaviour in which the second-law trend need not
hold locally in time \cite{ScharnhorstWolpertRovelli2024BoltzmannBridges}.
If \textit{ontic} understanding of randomness is adopted---i.e.\ assuming
that the effective stochastic terms represent physically real disturbances
rather than mere ignorance---this leads naturally to two classes of
behaviour in bridge-type settings: \textit{systemic} evolution,
characterised by non-decreasing entropy, decoherence-dominated dynamics and
stochastic processes evolving forward in time, and \textit{antisystemic}
evolution, characterised by non-increasing entropy, recoherence-dominated
dynamics and stochastic processes evolving backward in time \cite%
{Klimenko2025TwoTypesTemporalSymmetry}. While existence of antisystems is
permitted by the laws of nature (as we currently understand them),
antisystems may or may not exist in reality.

These preceding considerations are largely formulation-independent
(classical or quantum) and do not, by themselves, rely on specifically
quantum notions. \textsl{Our aim here is to express these ideas directly in
quantum-mechanical terms, using the formalism of Lindbladian master
equations \cite%
{Lindblad1976,Breuer2007book,WisemanMilburn2010QuantumMeasurementControl,LindbladIntro2020}
together with quantum-stochastic formulations, such as\ stochastic Schr\"{o}%
dinger equation \cite%
{GisinPercival1992QSDOpen,DephasingHAM2021,StochasticHamiltonians2026}}. If
two-time conditions are used then, conceptually, this has points of contact
with (i) the two-state vector formalism (TSVF) of pre- and post-selected
quantum ensembles \cite%
{AharonovBergmannLebowitz1964TimeSymmetry,Vaidman2007TwoStateVectorFormalism}%
, and (ii) quantum Schr\"{o}dinger bridges (QSB), which characterise
time-symmetric ensembles of quantum trajectories consistent with prescribed
endpoint density matrices \cite%
{MovillaMiangolarraSabbaghGeorgiou2025QuantumSchrodingerBridges}. However,
our emphasis is different: we retain an explicitly stochastic-realistic
component (random disturbances as physically operative) and use the
resulting systemic/antisystemic dichotomy to formulate alternative
fundamental symmetry properties for this component: CP- vs CPT-invariant.
This yields distinct predictions for decoherence and recoherence effects,
which are specifically considered in this work without addressing full
thermalisation. These predictions appear experimentally testable, in
particular via suitably chosen diffractive dissociations in proton--proton
and proton--antiproton collisions. \textsl{While quantum decoherence may be
viewed as a fundamental process underlying the thermodynamic arrow of time,
this work focuses exclusively on non-unitary dephasing effects and does not
address the broader thermodynamic, kinetic, or cosmological implications 
\cite{Klimenko2026Antisystems}. }

\textsl{The principal question examined in this work is, therefore, whether
non-unitary components associated with decoherence or recoherence, and
represented by Lindblad-type equations, are essential for the description of
diffractive dissociations. Section 2 introduces the relevant Lindblad models
for stochastic realism and double temporal conditioning. In line with the
perspective of stochastic realism, the intrinsic (Section 2.1) and
environmental (Section 2.2) approaches to the Lindblad equation are
interpreted as related but distinct realistic physical processes, rather
than as different models of the same physical process. Section 2.3 points to
a general trend that quantum non-resonant transitions tend to be amplified
by decoherence. Section 3 examines the symmetry and invariance properties of
non-unitary systems governed by Lindblad dephasing equations and suggests
two fundamental alternatives: CP invariance and CPT invariance of such
systems. The relevant covariant matrix forms of the Hamiltonian and Lindblad
operators are discussed in the Appendices. }

\textsl{Sections 4--6 are devoted to determining whether non-unitary terms
are essential for the description of diffractive dissociations. Section 4
analyses the effect of decoherence on diffractive interactions, with
particular emphasis on the triple-Pomeron vertex, which plays a key role in
diffractive dissociations, and derives a corresponding decoherence factor.
Section 5 uses the major published diffractive data from the principal
collaborations (ISR, UA4, UA5, CDF, D0, ALICE, and E710) to determine the
best experimental estimate of the decoherence factor for single diffraction
(Section 5.2) and double diffraction (Section 5.3). The possibility of
alternative explanations is examined in Section 6. The conclusions are
presented in Section 7.}

\section{Modelling of decoherence and stochastic realism}

There are two main philosophical approaches to \textit{randomness}: \textit{%
epistemic}, where randomness reflects incomplete knowledge and the need for
coarse-graining, and \textit{ontic}, where randomness is assumed to be
physically real and present in the world as a genuine process (i.e.\ \textit{%
stochastic realism}). The ontic and epistemic perspectives do not mirror
each other. The ontic view readily accommodates epistemic uncertainty
(ignorance and finite resolution remain even if randomness is fundamental),
whereas the epistemic view tends to regard ontic randomness as either
irrelevant to explanation or, in stronger forms, non-existent.

A standard illustration of epistemic randomness is tossing a coin: if we
knew the exact initial conditions and all relevant parameters, we could in
principle predict the outcome, but we typically do not, so the
result---heads or tails---appears random. By contrast, a ``quantum coin'',
such as a measurement of a spin prepared in a superposition $\propto \bigl(%
\left|\uparrow\right\rangle+\left|\downarrow\right\rangle\bigr)$, yielding
one of the outcomes $\left|\uparrow\right\rangle$ or $\left|\downarrow\right%
\rangle$ with equal probability, is often taken to illustrate ontic
randomness: within standard quantum mechanics, no additional information
about the prepared state allows one to predict the individual outcome. These
examples are illustrations rather than proofs. Philosophical positions such
as ``ontic'' versus ``epistemic'' are not, by themselves, experimentally
provable, because they are partly interpretive claims about what probability
means. What can be tested are physical models that embody these stances: as
long as models make different empirical predictions, they can be supported
or ruled out by data.

A stochastic form of the Schr\"{o}dinger equation can be written \cite%
{GisinPercival1992QSDOpen,WisemanMilburn2010QuantumMeasurementControl,DephasingHAM2021,StochasticHamiltonians2026}
as 
\begin{equation}
i\hbar \,d\left\vert \Psi \right\rangle ={H}\left\vert \Psi \right\rangle
\,dt+\sum_{j}{L}_{j}\left\vert \Psi \right\rangle \circ dW_{j},
\label{eq:stoch_schrod}
\end{equation}%
where ${H}={H}^{\dagger }$ is a conventional Hermitian Hamiltonian operator
(non-relativistic or relativistic as appropriate), ${L}_{j}={L}_{j}^{\dagger
}$ are Hermitian dephasing operators, and $W_{j}(t)$ are independent real
Wiener processes. The symbol "$\circ $" denotes the Stratonovich
interpretation, which is often adopted in order to preserve the usual rules
of calculus and to facilitate a time-symmetric formulation. In particular,
Wiener processes understood as running forward from past conditions or
running backward from future conditions have identical increment statistics,
so using either as the driver $W_{j}(t)$ in (\ref{eq:stoch_schrod}) leads to
the same class of noise realisations. By contrast, the Ito formulation is
not time-reversal covariant: reversing time in an Ito stochastic equation
requires more care because it induces additional drift terms and is
therefore not well suited to represent forward- and backward-running
processes simultaneously.

Equation~(\ref{eq:stoch_schrod}) is a quantum version of the general
stochastic equation~(A3) in \cite{Klimenko2025TwoTypesTemporalSymmetry},
which combines the perspective of stochastic realism with dual temporal
conditioning. It is straightforward to verify that the quantum form is
consistent with the odd-symmetry constraints (15) in \cite%
{Klimenko2025TwoTypesTemporalSymmetry}. Briefly, upon separating real and
imaginary parts (e.g. writing $\Psi =u+iv$ in a chosen basis and recasting
the dynamics for the real vector $(u,v)$), the resulting drift and diffusion
fields are linear in $(u,v)$, and their divergences reduce to traces of the
corresponding coefficient matrices. Hermiticity of ${H}$ and ${L}_{j}$
implies that their imaginary parts are antisymmetric (and hence trace-free),
so that $\func{Tr}\,\func{Im}\,{H}=0$ and $\func{Tr}\,\func{Im}\,{L}_{j}=0$,
which is the required condition of non-divergence. Compared with the
generalised Pauli master equation \cite{Klimenko2016}, the present framework
captures the progressive accumulation of decoherence induced by the
stochastic perturbation and is therefore suitable for pure states, maximally
mixed states, and any intermediate state.

\textsl{The connection between stochastic Schr\"{o}dinger equations and
Lindblad-type models is well known \cite{Breuer2007book} and is explored
further below.}

\subsection{The dephasing Lindblad equation}

Equation (\ref{eq:stoch_schrod}) can be easily transformed into an equation
for the density matrix $\rho $ --- stochastic version of the von Neumann's
equation%
\begin{equation}
i\hbar \,d\rho =\left[ {H},\rho \right] \,dt+\sum_{j}\left[ {L}_{j},\rho %
\right] \circ dW_{j},\ \ \ \ \ \rho =\sum_{i}\left\vert \Psi
_{i}\right\rangle \left\langle \Psi _{i}\right\vert  \label{eq:stoch_rho}
\end{equation}%
Our goal is to obtain the governing equations for the averages. This needs
to take into account the defining property of the Stratonovich stochastic
integral, namely that coefficients are evaluated at the midpoint of each
time step, creating correlations between the diffusion coefficient and $%
dW_{j}$. Upon ensemble averaging over realisations of $W_{j}(t)$ (denoted by
an overtilde), the midpoint terms proportional to $dW_{k}dW_{j}$ produce the
following $O(dt)$ contributions: 
\begin{equation}
\widetilde{\mathstrut {L}_{j}\left\vert \Psi \right\rangle \circ dW_{j}}={L}%
_{j}\left\vert \widetilde{\Psi }\right\rangle \widetilde{dW_{j}}+\frac{{L}%
_{j}}{2i\hbar }\sum_{k}{L}_{k}\left\vert \widetilde{\Psi }\right\rangle 
\widetilde{dW_{k}dW_{j}}=\frac{{L}_{j}^{2}\left\vert \widetilde{\Psi }%
\right\rangle }{2i\hbar }\gamma _{j}dt
\end{equation}%
\begin{equation}
\widetilde{\mathstrut \left[ {L}_{j},\rho \right] \circ dW_{j}}=\left[ {L}%
_{j},\tilde{\rho}\right] \widetilde{dW_{j}}+\frac{1}{2i\hbar }\sum_{k}\left[ 
{L}_{j},\left[ {L}_{k},\tilde{\rho}\right] \right] \widetilde{dW_{k}dW_{j}}=%
\frac{\left[ {L}_{j},\left[ {L}_{j},\tilde{\rho}\right] \right] }{2i\hbar }%
\gamma _{j}dt
\end{equation}%
where we use the following identities 
\begin{equation}
\widetilde{dW_{j}}=0,\ \ \ \widetilde{dW_{k}dW_{j}}=\delta _{kj}\left\vert
dt\right\vert
\end{equation}%
and, implicitly, a Dyson-like series to evaluate half-step increments for $%
\left\vert \Psi \right\rangle $ and $\rho $. Since under bridge conditions
the effective driving processes can be run both forward and backward in time
the absolute value $\left\vert dt\right\vert $ indicates universal
positiveness of the term $\widetilde{dW_{j}dW_{j}}$. Therefore, $\widetilde{%
dW_{j}dW_{j}}=\gamma _{j}dt$ where $\gamma _{j}=1$ when the stochastic
process $W_{j}(t)$ runs forward in time or $\gamma _{j}=-1$ when the
stochastic process $W_{j}(t)$ runs backward in time. While microscopically
equivalent, the forward and backward processes have radically different
average properties, which are associated with forward-time and reverse-time
parabolicities and these properties tend to persist indefinitely.

Averaging of equation (\ref{eq:stoch_schrod}), therefore, leads to 
\begin{equation}
i\hbar \frac{d\left\vert \widetilde{\Psi }\right\rangle }{dt}={H}\left\vert 
\widetilde{\Psi }\right\rangle \,-\frac{i}{2\hbar }\sum_{j}\gamma _{j}{L}%
_{j}^{2}\left\vert \widetilde{\Psi }\right\rangle ,  \label{eq:psi}
\end{equation}%
while averaging of equation (\ref{eq:stoch_rho}) yields 
\begin{equation}
i\hbar \frac{d\tilde{\rho}}{dt}=\left[ {H},\tilde{\rho}\right] -\mathcal{D(}%
\tilde{\rho}),\ \ \ \ \tilde{\rho}=\sum_{i}\widetilde{\left\vert \Psi
_{i}\right\rangle \left\langle \Psi _{i}\right\vert }  \label{eq:rho}
\end{equation}%
\begin{equation}
\mathcal{D(}\tilde{\rho})=\frac{i}{2\hbar }\sum_{j}\gamma _{j}\left[ {L}_{j},%
\left[ {L}_{j},\tilde{\rho}\right] \right] =\frac{i}{2\hbar }\sum_{j}\gamma
_{j}\left( \left\{ {L}_{j}^{2},\tilde{\rho}\right\} -2{L}_{j}\tilde{\rho}\,{L%
}_{j}\right)  \label{eq:Lind}
\end{equation}%
is, for positive $\gamma _{j}$, a conventional Lindblad form of the
dephasing operator $\mathcal{D}\tilde{\rho}=\mathcal{D(}\tilde{\rho})$
evaluated under Hermitian restrictions ${L}_{j}={L}_{j}^{\dagger }$.
Calligraphic letters (e.g. $\mathcal{D}$) denote superoperators or linear
maps that act in the Liouvillian space of quantum operators. In addition to $%
\gamma _{j}=\pm 1$, one may set $\gamma _{j}=0$ to omit the corresponding
operator $L_{j}$, or allow $\gamma _{j}$ to take arbitrary real values so
that $\left\vert \gamma _{j}\right\vert $ reflects the magnitude of the $%
L_{j}$ term.\ While equations~(\ref{eq:rho})--(\ref{eq:Lind}) coincide with
the Lindblad dephasing equation for $\gamma _{j}>0$, which describes
decoherence processes and entropy increase in a very general Markovian
setting. As the dephasing operator is unital $\mathcal{D(}I)=0$,
non-negative $\gamma _{j}\geq 0$ ensure that von Neumann entropy is
non-decreasing in time \cite{Abe2017}, while\ negative $\gamma _{j}<0$
correspond to recoherence and to entropy-decreasing behaviour. Unlike
bi-directional Pauli master equation \cite{Klimenko2016}, equation (\ref%
{eq:rho}) preserves the quantum-mechanical structure based on density
matrices and can describe progressively developing decoherence. \textsl{%
While conventional Lindblad equation can be interpreted as
quantum-mechanical version of the conventional Fokker--Planck equation \cite%
{Breuer2007book},}\ this equation with positive and negative $\gamma _{j}$
may be viewed as a quantum-mechanical analogue of the bi-directional
(odd-symmetric) Fokker--Planck equation \cite%
{Klimenko2025TwoTypesTemporalSymmetry}.\ This bi-directional duality is
associated with dual temporal boundary conditions, which have been discussed
in a number of publications \cite%
{Schulman1997TimesArrowsBook,Tamm2021,ScharnhorstWolpertRovelli2024BoltzmannBridges}%
. Our stochastic-realist perspective tends to emphasise the role of
physically real stochastic processes that can, at least in principle, be
running forward or backward in time. Note that practical forward-time
applications of reverse-time (inverse-parabolic) models---equivalently,
generators with negative dephasing rates $\gamma _{i}<0$ in a
time-homogeneous Lindblad-type form---are limited. Such dynamics is
typically ill-posed, in the sense that small perturbations can amplify
rapidly, and it may reach the boundaries of physically admissible behaviour
(e.g. loss of complete positivity). Accordingly, the significance of these
equations is primarily conceptual rather than applied.

\textsl{The Lindblad equation with }$\gamma _{j}\geq 0$\textsl{\ is known to
preserve positivity and trace \cite%
{Lindblad1976,Breuer2007book,LindbladIntro2020}. Formally running this
evolution backward in time would still preserve positivity and trace, even
though the corresponding coefficients would satisfy }$\gamma _{j}\leq 0$%
\textsl{\ in the time-reversed frame. However, once the evolution is
expressed in this reversed-time form, positivity and trace preservation are
not guaranteed universally, since the process must be terminated whenever
the density matrix reaches a degenerate state. The need for such termination
conditions is standard in antisystemic and reverse-time models \cite%
{Klimenko2017KineticsCPT}. In practice, negative }$\gamma _{j}$\textsl{\ may
indicate a reduction of the overall decoherence rate rather than sustained
recoherence, but such values should be allowed for completeness when
extending the forward-time semigroup to a bi-directional group.}

The Fokker-Planck equations are formulated for probability distributions.
Although the bi-directional Lindblad dephasing equation~(\ref{eq:rho})--(\ref%
{eq:Lind}) is written for the ensemble-averaged density $\tilde{\rho}$
rather than for an explicit probability density function, it is sufficient
for our purposes. Indeed, the stochastic Schr\"{o}dinger equation~(\ref%
{eq:stoch_schrod}) is linear and therefore admits closed evolution equations
for its moments. In this context, equation~(\ref{eq:psi}) describes the
first moment, while equation~(\ref{eq:rho}) governs the second. An
additional advantage of the quantum-mechanical formulation is the
convenience of analysing the covariant properties of the bi-directional
Lindblad equation.

\bigskip

The number of linearly independent operators ${L}_{j}$ on an $n_{\rho }$%
-dimensional Hilbert space is $n_{\rho }^{2}-1$ (excluding the identity
operator ${L}_{j}={I}$, which does not affect $\mathcal{D(}\tilde{\rho})$.
Accordingly, a completely general Lindblad generator may, in principle,
involve up to $n_{\rho }^{2}-1$ independent noise channels. If the
decoherence basis is fixed (usually to the system's eigenstates) and the
dynamics is restricted to dephasing in that basis \cite%
{SchirmerSolomon2004,OiSchirmer2012,SN-AS2021,Mercurio_2023}, then in this
basis $\left\vert 1\right\rangle ,$ $\left\vert 2\right\rangle ,$ ..., $%
\left\vert n_{\rho }\right\rangle $ operators ${L}_{j}$ are diagonal and
equation (\ref{eq:rho})--(\ref{eq:Lind}) can be modified to take the form 
\begin{equation}
{L}_{j}=\sum_{k}a_{j}^{k}\left\vert k\right\rangle \left\langle k\right\vert
,\ \ \ \ \left\langle k\right\vert \mathcal{D(}\tilde{\rho})\left\vert
i\right\rangle =i{\Gamma }_{ki}\tilde{\rho}_{ki},\ \ \ \ \ {\Gamma }_{ki}=%
\frac{1}{2\hbar }\sum_{j}\gamma _{j}\left( a_{j}^{k}-a_{j}^{i}\right) ^{2}
\label{eq:Lam_stoc}
\end{equation}%
The number of independent operators ${L}_{j}$ is $n_{\rho }-1$ in this case (%
$n_{\rho }$ independent vectors $\mathbf{a}_{j}$ excluding the identity
operator ${I}$ with $\mathbf{a}_{n_{\rho }}=\mathbf{1}$), which is much
smaller than $n_{\rho }^{2}-1.$ In the bi-directional formulation, the
number of ${L}_{j}$ can be increased to account for both decoherence $\gamma
_{j}>0$ and recoherence $\gamma _{j}<0$. If environmental and intrinsic
mechanisms are modelled as distinct contributions, the number of Lindblad
operators under consideration may \ \ \ increase further. \textsl{In the
case of pure dephasing, which does not induce transitions between energy
eigenstates, the dephasing basis coincides with the energy eigenbasis; the
details are discussed in the Appendices.}

\subsection{Note on bath dephasing}

\label{subsec:bath_dephasing}

The stochastic terms in (\ref{eq:stoch_schrod}) and (\ref{eq:stoch_rho}) may
be interpreted either as genuinely random, intrinsic disturbances (\textit{%
ontic randomness}) or as an effective description of interactions with an
environment whose microscopic degrees of freedom are not fully known or
tracked (\textit{epistemic randomness}). The cumulative effect of numerous
small, uncontrolled perturbations is often difficult to predict and, as
demonstrated in \cite{Breuer2007book,LindbladIntro2020} and outlined below,
can lead (after tracing out the bath) to reduced dynamics that is
statistically similar to that generated by explicitly stochastic terms.

We employ the conventional system--bath decomposition 
\begin{equation}
{H}={H}_{0}+{H}_{\text{int}},\ \ \ {H}_{0}={H}_{S}\otimes {I}_{B}+{I}%
_{S}\otimes {H}_{B},\ \ \ {H}_{\text{int}}=\sum_{\beta }{L}_{\beta }\otimes {%
V}_{\beta },  \label{eq:HSB_decomposition}
\end{equation}%
where ${L}_{\beta }$ act on the system Hilbert space and ${V}_{\beta }$ act
on the bath as a part of the system--bath interaction Hamiltonian ${H}_{%
\text{int}}$. Working in the interaction picture with respect to ${H}_{0}$,
define 
\begin{equation}
\rho _{I}(t)\equiv {U}_{0}^{\dagger }(t)\rho _{SB}(t){U}_{0}(t),\qquad {H}%
_{I}(t)\equiv {U}_{0}^{\dagger }(t){H}_{\text{int}}{U}_{0}(t),\qquad {U}%
_{0}(t)=e^{-\frac{i}{\hbar }{H}_{0}t}.
\end{equation}%
---the interaction-picture quantities $\rho _{I}$ and ${H}_{I}$ based on the
unitary transformation ${U}_{0}(t).$ The von Neumann equation for the joint
system--bath state becomes 
\begin{equation}
i\hbar \frac{d\rho _{I}(t)}{dt}=\Big[{H}_{I}(t),\rho _{I}(t)\Big].
\label{eq:lvn_intpic}
\end{equation}

\bigskip

Dyson-like (iterative) integration of (\ref{eq:lvn_intpic}) yields the exact
identity 
\begin{equation}
i\hbar \frac{d\rho _{I}(t)}{dt} =\Big[ {H}_{I}(t),\rho _{I}(0)\Big] +\frac{1%
}{i\hbar }\int_{0}^{t}dt_{1}\, \Big[ {H}_{I}(t),\left[ {H}_{I}(t_{1}),\rho
_{I}(t_{1})\right] \Big].  \label{eq:Dyson}
\end{equation}
(The second term is second order in ${H}_{\text{int}}$ and will generate the
Redfield kernel after tracing out the bath.)

The environmental degrees of freedom are then traced out: 
\begin{equation}
i\hbar \frac{d\tilde{\rho}_{I}(t)}{dt} =\mathrm{Tr}_{B}\Big[ {H}_{I}(t),\rho
_{I}(0)\Big] +\frac{1}{i\hbar }\int_{0}^{t}dt_{1}\, \mathrm{Tr}_{B}\Big[ {H}%
_{I}(t),\left[ {H}_{I}(t_{1}),\rho _{I}(t_{1})\right] \Big],
\end{equation}
where $\tilde{\rho}_{I}\equiv \mathrm{Tr}_{B}\left( \rho _{I}\right)$ is the
reduced density matrix of the system in the interaction picture. The first
(formally first-order) term $\mathrm{Tr}_{B}\big[ {H}_{I}(t),\rho _{I}(0)%
\big]$ is conventionally neglected by assuming $\mathrm{Tr}_{B}( {V}%
_{j}\rho_{B})=0$ (or, equivalently, that any mean-field contribution can be
absorbed into the system Hamiltonian).

The principal assumption made in this context is the \textit{Born
approximation}%
\begin{equation}
\rho _{SB}(t)\approx \tilde{\rho}(t)\otimes \rho _{B}  \label{eq:born_approx}
\end{equation}%
where the bath density matrix is presumed to be in its steady-state $[ {H}%
_{B},\rho _{B}]=0$. The Born approximation declares that the system and the
bath are effectively decorrelated during interactions, and correlations
appear as a result of these interactions, as specified by the double
commutator in (\ref{eq:Dyson}). In conjunction with forward-time
integration, the Born approximation becomes strongly time-directional. This
is analogous to Boltzmann's hypothesis of molecular chaos: it presumes an
absence of correlations before, but not after, interactions.

This yields a Redfield-type equation for $\tilde{\rho}_{I}(t)$. For clarity,
we display the single-channel case ${H}_{\mathrm{int}}={L}\otimes {V}$ (or ${%
H}_{I}(t)={L}_{I}(t)\otimes {V}_{I}(t)$), for which one obtains 
\begin{equation}
\frac{d\tilde{\rho}_{I}(t)}{dt}=-\frac{1}{\hbar ^{2}}\int_{0}^{t}d\tau \,%
\Big(C(\tau )\,[{L}_{I}(t),{L}_{I}(t-\tau )\tilde{\rho}_{I}(t-\tau
)]-C(-\tau )\,[{L}_{I}(t),\tilde{\rho}_{I}(t-\tau ){L}_{I}(t-\tau )]\Big),
\label{eq:redfield}
\end{equation}%
where $\tau =t-t_{1}$ and the bath correlation function is 
\begin{equation}
C(\tau )\equiv \mathrm{Tr}_{B}\!\left( e^{\frac{i}{\hbar }{H}_{B}\tau }\,{V}%
\,e^{-\frac{i}{\hbar }{H}_{B}\tau }\,{V}\,\rho _{B}\right) .
\label{eq:bath_corr}
\end{equation}%
For a stationary bath and Hermitian ${V}$, this correlator satisfies $%
C(-\tau )=C^{\ast }(\tau )$ so that $\mathrm{Re}\,C(\tau )$ is even and $%
\mathrm{Im}\,C(\tau )$ is odd.

Assuming a short bath correlation time (Markov limit), the upper limit in (%
\ref{eq:redfield}) can be extended to $\infty $, and over the range $\tau
\lesssim \tau _{C}$ one may approximate $\tilde{\rho}_{I}(t-\tau )\approx 
\tilde{\rho}_{I}(t)$. We then define 
\begin{equation}
\frac{\gamma ^{B}}{2}\equiv \int_{0}^{\infty }d\tau \,\mathrm{Re}\,C(\tau
),\qquad \frac{\eta ^{B}}{2}\equiv \int_{0}^{\infty }d\tau \,\mathrm{Im}%
\,C(\tau ).  \label{eq:gamma_eta}
\end{equation}%
In the pure-dephasing case (so that ${L}_{I}(t)={L}$), one may return to the
Schr\"{o}dinger picture and reduce (\ref{eq:redfield}) to 
\begin{equation}
i\hbar \frac{d\tilde{\rho}}{dt}=\left[ {H}_{S}+\frac{\eta ^{B}}{2\hbar }\,{L}%
^{2},\tilde{\rho}\right] -\underset{\mathcal{D}_{B}\mathcal{(}\tilde{\rho})}{%
\underbrace{\frac{i}{2\hbar }\gamma ^{B}\left[ {L},\left[ {L},\tilde{\rho}%
\right] \right] }}.  \label{eq:lindblad_bath}
\end{equation}%
The first term proportional to $\eta _{B}$ is the usual Lamb-shift
contribution, which is conventionally neglected or absorbed into the system
Hamiltonian. The remaining term is of the same double-commutator dephasing
form as in (\ref{eq:Lind}), with a bath-induced coefficient $\gamma _{B}$,
which is predominantly positive under the stated assumptions (equivalently,
it is determined by a non-negative noise spectrum). We take into account
multiple interactions in $H_{\text{int}}$ given in (\ref%
{eq:HSB_decomposition}) and sum up stochastic and bath-interaction terms
assuming pure dephasing 
\begin{equation}
L_{\beta }^{B}=\sum_{k}b_{\beta }^{k}\left\vert k\right\rangle \left\langle
k\right\vert ,\ \ \ \left\langle k\right\vert \mathcal{D}_{B}\mathcal{(}%
\tilde{\rho})\left\vert i\right\rangle =i{\Gamma }_{ki}^{B}\tilde{\rho}%
_{ki},\ \ \ {\Gamma }_{ki}^{B}=\frac{1}{2\hbar }\sum_{\beta }\gamma _{\beta
}^{B}\left( b_{\beta }^{k}-b_{\beta }^{i}\right) ^{2}  \label{eq:Lam_bath}
\end{equation}

\textsl{Considering dual temporal boundary conditions, the non-negativity of 
}$\gamma _{j}^{B}$\textsl{\ is not assured. Indeed, let us consider the
time-symmetric Hamiltonians in (\ref{eq:HSB_decomposition}) for a
time-invariant quantum system immersed in an antibath, i.e.\ a bath subject
to future temporal conditions and equilibrated by backward-time evolution.
For such an antibath, induced dependencies appear before interactions and
disappear after interactions; its behaviour is therefore essentially
retrocausal. The microscopic system--bath problem may still be formulated
with the same Hamiltonian, but the reduced description now requires advanced
rather than retarded closing assumptions. Equivalently, the problem can be
solved backward in time under the same approximations as those used for an
ordinary bath in forward time. When rewritten in the forward-time
description, this reverses the sign of the effective dephasing coefficient,
so that the corresponding values of }$\gamma _{j}^{B}$\textsl{\ become
negative. If the system is coupled to several non-interacting baths and
antibaths, some of the }$\gamma _{j}^{B}$\textsl{\ may therefore be positive
and others negative.}

\textsl{While we still assume short correlations --- delta-correlated
processes give rise to Markovian properties \cite{Klyatskin1991} --- such
processes may nevertheless correspond, in the forward-time representation,
to either forward or inverse diffusivity. Effective negative values of }$%
\gamma ^{B}$\textsl{\ may also arise from long correlations and from
violations of Markovianity, since quantum information can flow in both
directions, from the system to the bath and back \cite{Hall2014,Breuer2016}.
At the same time, Markovianity and non-negative }$\gamma ^{B}$\textsl{\ (or
forward parabolicity of the Fokker--Planck equation) are not uniquely linked 
\cite{Breuer2016,Tarasov2021}. Under observable real-world conditions,
however, thermodynamic antisystems are not observed and, assuming
Markovianity and forward causality, }$\gamma _{j}^{B}$\textsl{\ should
remain non-negative.}

While environmental disturbances can provide a microscopic justification for
introducing stochastic contributions into the von Neumann equation, our
stochastic-realist perspective treats both intrinsic and environmental
mechanisms of decoherence as real and therefore as jointly affecting quantum
systems. In some cases, environmental and intrinsic sources of decoherence
may lead to distinct observable signatures \cite{SN-AS2021} or to intrinsic
Hamiltonian differences that cannot be removed by a change of representation
(i.e. a \textquotedblleft Hamiltonian obstruction\textquotedblright\ in the
sense of \cite{Fagnola2019MarkovianDephasing}). It seems, however, that when
both sources of decoherence are present, any such differences are likely to
be masked by their combined effect. If stochastic realism is adopted, and if
these disturbances are taken to represent forward-time stochastic processes
(i.e. processes generated from initial conditions), they will lead to a
progressive loss of system--environment correlations in forward time. This
is precisely the condition required for using the Born closure together with
forward-time integration: stochastic realism thereby fixes the arrow of time
for environmental interactions through the forward development of random
disturbances. Finally, to evaluate the total dephasing effect, we combine
the stochastic and bath-induced contributions 
\begin{equation}
\left\langle k\right\vert \mathcal{D(}\tilde{\rho})\left\vert i\right\rangle
_{\text{tot}}=i{\Gamma }_{ki}^{\text{tot}}\tilde{\rho}_{ki},\ \ \ {\Gamma }%
_{ki}^{\text{tot}}={\Gamma }_{ki}+{\Gamma }_{ki}^{B}\   \label{eq:Lam_tot}
\end{equation}%
where ${\Gamma }_{ki}$ is defined in (\ref{eq:Lam_stoc}).

\subsection{A minimal model for diffractive dissociation: Lindblad--Dyson
formulation}

\label{subsec:sd_twochannel_lindblad}

Diffractive dissociations and inelastic reactions are examples of processes
that often involve decoherence in one form or another. When decoherence
becomes an important constituent of the effective dynamics, it is natural to
use Lindblad-type master equations, which incorporate such non-unitary
effects in a controlled way, rather than relying solely on the unitary Schr%
\"{o}dinger or von Neumann equations for a closed system. Conversely, if
decoherence is only secondary for the observable of interest, a framework
that includes it should show that the corresponding corrections are small
and may be neglected. Since wave-function and amplitude methods do not
readily accommodate decoherence without moving to a density-operator
description and tracing over unobserved degrees of freedom, this motivates
the use of density-matrix formulations, even if they can be less convenient
in other respects.

This subsection illustrates the effect of decoherence on the transition of a
quantum system from state~1 to state~2, characterised by the corresponding
change of the density operator, $\tilde{\rho}(\text{state 1})\rightarrow 
\tilde{\rho}(\text{state 2})$. State~1 may represent an incoming
(elastic-like) channel, while state~2 represents an excited channel, e.g.\
dissociative reaction that subsequently produces $X$, a multi-particle
hadronised final state. We take the initial condition $\tilde{\rho}_{11}=1$,
with all other components of $\tilde{\rho}$ vanishing. After the
interaction, a nonzero population $\tilde{\rho}_{22}>0$ indicates a finite
transition probability, which is directly related (and, with appropriate
kinematic binning and normalisation, proportional) to the cross-section for
the corresponding collision channel.

Consistently with the previous analysis, the dephasing Lindblad equation can
be written in the form 
\begin{equation}
\frac{d\tilde{\rho}}{dt}=-\frac{i}{\hbar }\,[H,\tilde{\rho}]-\frac{\gamma }{%
2\hbar ^{2}}\,[L,[L,\tilde{\rho}]].  \label{eq:Lind2}
\end{equation}%
assuming here that $\gamma \geq 0$ is the dephasing intensity and only one
Lindblad operator $L$ is present in $\mathcal{D}$.

We split the Hamiltonian into a \textquotedblleft main\textquotedblright\
part and a weak off-diagonal coupling $H=H_{0}+gH_{1}$ and restrict the
dynamics to two orthogonal eigenstates of $H_{0}$: the initial state $%
|1\rangle $ and the final state $|2\rangle $. The dephasing operator $L$ is
taken diagonal in this basis (pure dephasing), so it does not induce direct
transitions; conversion between the two states is driven solely by the
off-diagonal coupling $H_{1}$. Without loss of generality we set 
\begin{equation}
H_{0}=\sum_{i=1}^{2}E_{i}\,|i\rangle \langle i|,\qquad H_{1}=\hbar \Big(%
|1\rangle \langle 2|+|2\rangle \langle 1|\Big)\equiv \hbar \sigma
_{x},\qquad L_{0}=\sum_{i=1}^{2}a_{i}\,|i\rangle \langle i|.  \label{eq:H0L}
\end{equation}%
where $\sigma _{x}$ denotes the Pauli $x$-matrix. We define the detuning and
the total dephasing rate 
\begin{equation}
\Delta \omega \overset{\text{{\tiny def}}}{=}\frac{E_{2}-E_{1}}{\hbar }%
,\qquad \Gamma ^{\text{tot}}=\Gamma _{0}\overset{\text{{\tiny def}}}{=}\frac{%
\gamma }{2\hbar ^{2}}(a_{1}-a_{2})^{2}.  \label{eq:DeltaGamma}
\end{equation}%
Equation (\ref{eq:Lind2}) is written as 
\begin{equation}
\frac{d\tilde{\rho}}{dt}=\mathcal{L}_{0}\tilde{\rho}+g\,\mathcal{L}_{1}%
\tilde{\rho}  \label{eq:L0L1}
\end{equation}%
where 
\begin{equation}
\mathcal{L}_{0}\tilde{\rho}\overset{\text{{\tiny def}}}{=}-\frac{i}{\hbar }%
[H_{0},\tilde{\rho}]-\frac{\gamma }{2\hbar ^{2}}[L_{0},[L_{0},\tilde{\rho}%
]],\qquad \mathcal{L}_{1}\tilde{\rho}\overset{\text{{\tiny def}}}{=}-\frac{i%
}{\hbar }[H_{1},\tilde{\rho}]=-\,i\,[\sigma _{x},\tilde{\rho}].
\label{eq:L0L1_defs}
\end{equation}%
We expand the solution in powers of the Born coupling $g$: 
\begin{equation}
\tilde{\rho}(t)=\rho ^{(0)}(t)+g\,\rho ^{(1)}(t)+g^{2}\rho ^{(2)}(t)+\cdots .
\label{eq:rho_series}
\end{equation}%
Substituting into (\ref{eq:L0L1}) and collecting powers of $g$ yields 
\begin{equation}
\frac{d\rho ^{(0)}}{dt}=\mathcal{L}_{0}\rho ^{(0)},\qquad \frac{d\rho
^{(i+1)}}{dt}=\mathcal{L}_{0}\rho ^{(i+1)}+\mathcal{L}_{1}\rho ^{(i)}\quad
(i\geq 0),  \label{eq:hierarchy}
\end{equation}%
with $\rho ^{(0)}(0)=\tilde{\rho}(0)$ and $\rho ^{(i)}(0)=0$ for $i>0$.
Integration gives the Lindbladian analogue of the Dyson series, which has
been introduced and used in recent publications \cite%
{Lind-Dyson2023,Lind-Dyson2025} 
\begin{align}
\rho ^{(0)}(t)& =e^{t\mathcal{L}_{0}}\,\rho (0), \\[0.04in]
\rho ^{(1)}(t)& =\int_{0}^{t}\!dt_{1}\;e^{(t-t_{1})\mathcal{L}_{0}}\,%
\mathcal{L}_{1}\,e^{t_{1}\mathcal{L}_{0}}\,\rho (0), \\[0.04in]
\rho ^{(2)}(t)& =\int_{0}^{t}\!dt_{1}\!\int_{0}^{t_{1}}\!dt_{2}\;e^{(t-t_{1})%
\mathcal{L}_{0}}\,\mathcal{L}_{1}\,e^{(t_{1}-t_{2})\mathcal{L}_{0}}\,%
\mathcal{L}_{1}\,e^{t_{2}\mathcal{L}_{0}}\,\rho (0).
\label{eq:lindblad_dyson}
\end{align}%
The exponential $e^{t\mathcal{L}_{0}}$ is a \textit{superoperator} (linear
map) propagating density matrices under the unperturbed generator $\mathcal{L%
}_{0}$. For the present two-state problem its action is explicit: 
\begin{equation}
\rho (\tau +t_{0})=e^{\tau \mathcal{L}_{0}}\rho (t_{0})=%
\begin{pmatrix}
\rho _{11}(t_{0}) & e^{-(\Gamma _{0}-i\Delta \omega )\tau }\,\rho
_{12}(t_{0}) \\[4pt] 
e^{-(\Gamma _{0}+i\Delta \omega )\tau }\,\rho _{21}(t_{0}) & \rho
_{22}(t_{0})%
\end{pmatrix}%
,  \label{eq:sd_L0_action}
\end{equation}%
i.e.\ $\mathcal{L}_{0}$ leaves populations unchanged while damping and
rotating the coherence.

With $\rho ^{(0)}(0)=\tilde{\rho}(0)=|1\rangle \langle 1|$ we have $\rho
^{(0)}(t)=|1\rangle \langle 1|$, and the first-order correction has only
off-diagonal entries: 
\begin{equation}
\rho _{11}^{(1)}(t)=\rho _{22}^{(1)}(t)=0,\qquad \rho _{12}^{(1)}(t)=\rho
_{21}^{(1)}(t)^{\ast }=i\,\frac{1-e^{(-\Gamma _{0}+i\Delta \omega )t}}{%
\Gamma _{0}-i\Delta \omega }\,,  \label{eq:sd_rho12_first}
\end{equation}%
where $\rho _{jk}=\langle j|\rho |k\rangle $.

\bigskip

The first non-zero $\rho _{22}$ appears at second order. Using (\ref%
{eq:hierarchy}) and the fact that $\mathcal{L}_{0}$ does not change
populations, one obtains (to order $g^{2}$ in the full density matrix) 
\begin{equation}
\,\frac{d\rho _{22}^{(2)}}{dt}=2\,\Phi _{t},\qquad \Phi
_{t}=\int_{0}^{t}\!dt_{2}\;e^{-\Gamma _{0}(t-t_{2})}\cos \!\big(\Delta
\omega (t-t_{2})\big).  \label{eq:sd_rho22_rate}
\end{equation}%
In the long-interaction-time limit $\Phi _{t}\rightarrow $ $\Phi $ as $%
t\rightarrow \infty ,$ we obtain an expression for the effective coupling
coefficient $g_{1\rightarrow 2}^{2}$%
\begin{equation}
\frac{d\tilde{\rho}_{22}}{dt}=2g_{1\rightarrow 2}^{2},\ \ g_{1\rightarrow
2}^{2}=g^{2}\Phi ,\ \ \ \Phi =\frac{\Gamma _{0}}{\Gamma _{0}^{2}+\Delta
\omega ^{2}}\overset{{\scriptscriptstyle\Gamma }_{0}{\ll \Delta \omega }}{%
\approx }\frac{\Gamma _{0}}{\Delta \omega ^{2}}=\frac{\Gamma ^{\text{tot}}}{%
\Delta \omega ^{2}}  \label{eq:sd_Phi_asymptote}
\end{equation}%
that accounts for decoherence. Thus a finite dephasing scale $\gamma >0$
enhances the transition, while strong dephasing reverses this trend. \textsl{%
The superscript "tot" is used to emphasise that }$\Gamma $\textsl{\
represents total (intrinsic+environmental) dephasing defined in (\ref%
{eq:Lam_tot}).} \textsl{If the dephasing intensity is changed from }$\Gamma
_{1}^{\text{tot}}$\textsl{\ to }$\Gamma _{2}^{\text{tot}}$\textsl{\ then its
effect on transition or reaction is given by the \textit{relative
decoherence factor} }$\phi \overset{\text{{\tiny def}}}{=}\Phi _{2}/\Phi
_{1}\approx \Gamma _{2}^{\text{tot}}/\Gamma _{1}^{\text{tot}}$.

The analysis is based on a Lindbladian version of the Dyson series \cite%
{Lind-Dyson2023,Lind-Dyson2025}. Although formally analogous to the
conventional expansion, it involves time integrals over superoperators and
therefore requires some care. We consider pure dephasing, with Lindblad
operators aligned with the eigenstates of the principal Hamiltonian so that
dephasing alone cannot drive a direct transition between the initial and the
final states. Nevertheless, in the present setting dephasing modifies the
effective conversion strength through the multiplicative decoherence factor $%
\Phi $---it broadens the transition channel and can assist non-resonant
transfer by suppressing coherent recurrences, although sufficiently strong
dephasing ultimately reduces the transition probability (quantum-Zeno
effect). This behaviour is commonly observed in quantum systems: for
example, non-resonant tunneling is enhanced and resonant tunneling is
suppressed by decoherence \cite{SN-AS2021}. This subsection illustrates that
decoherence can enhance a broad class of quantum transformations.

\section{Symmetries and the arrow of time in reduced quantum dynamics}

Physical symmetries are fundamental to quantum theory~\cite{Symmetry1972}
and constrain not only unitary dynamics but also decohering reduced dynamics~%
\cite{K-PhysA,Klimenko2016}. By Wigner's theorem, a physical symmetry is
represented on Hilbert space either by a unitary operator (denoted here by $%
\mathrm{U}$) or by an antiunitary operator (denoted by $\Theta $). Charge
conjugation C, parity P, and the combined transformation CP are examples of
unitary symmetries, whereas time reversal T and the combined transformation
CPT are antiunitary symmetries. Any antiunitary operator can be written as $%
\Theta =\mathrm{U}\mathrm{K}$, where $\mathrm{U}$ is unitary and $\mathrm{K}$
denotes complex conjugation in a chosen basis.

\subsection{Symmetry operators and superoperator covariance}

Invariance of an operator $H$ under a symmetry $\mathrm{S}$ means that the
transformed and original operators are equivalent, $\mathrm{S}H\mathrm{S}%
^{-1}=H$, which implies $\mathrm{S}H|\psi\rangle=H\mathrm{S}|\psi\rangle$
for any state $|\psi\rangle$. This may be written as 
\begin{equation}
\mathrm{S}H\mathrm{S}^{-1}=H\text{ \ or \ }\mathrm{S}H=H\mathrm{S}\text{ }
\label{sym1}
\end{equation}

According to modern physics, the laws governing fundamental interactions are
expected to be invariant under the combined \textrm{CPT} transformation
(simultaneous charge conjugation, parity inversion and time reversal), and
many interactions are also approximately \textrm{CP}-invariant. Although 
\textrm{CP} violation is known to occur in certain processes, such cases are
relatively exceptional. In this work we therefore assume that all
Hamiltonians $H$ under consideration are both \textrm{CP}- and \textrm{CPT}%
-invariant: 
\begin{equation}
\mathrm{U}_{\text{CP}}H\mathrm{U}_{\text{CP}}^{-1}=\mathrm{U}_{\text{CP}}H%
\mathrm{U}_{\text{CP}}^{\dag }=H,\text{ \ \ }{\Theta }_{\text{CPT}}H{\Theta }%
_{\text{CPT}}^{-1}=H
\end{equation}%
Under this assumption, the Hamiltonians are also \textrm{T}-invariant, since
the CPT operator is understood as ${\Theta }_{\text{CPT}}=\mathrm{U}_{\text{%
CP}}{\Theta }_{\text{T}}$ and combining CP- and CPT-invariance yields $%
\Theta _{\text{T}}H\Theta _{\text{T}}^{-1}=H$.

In both the stochastic and microscopic bath pictures it is natural to treat
the dephasing operators as time-reversal covariant. We follow this
convention and do not wish to artificially introduce temporal asymmetry;
accordingly, we assume all dephasing operators to be \textrm{T}-invariant in
the sense of (\ref{symTHL}). These relations can be written in the form 
\begin{equation}
{\Theta }_{\text{T}}H{\Theta }_{\text{T}}^{-1}=H,\ \ \Theta _{\text{T}}L_{j}{%
\Theta }_{\text{T}}^{-1}=\pm L_{j},\ \ {\Theta }_{\text{T}}L_{\beta }^{B}{%
\Theta }_{\text{T}}^{-1}=\pm L_{\beta }^{B}  \label{symTHL}
\end{equation}%
The sign in front of the operators $L$ does not affect pure dephasing
(because the dephasing term depends on $L$ quadratically) and generally does
not need to be tracked explicitly. \textsl{The structure of Lindblad
operators }$L$\textsl{\ tends to be block-diagonal under these
assumptions---the details are considered in the Appendices. }

\bigskip

The conventional operator symmetries are now generalised to superoperators.
The symmetry condition (\ref{sym1}) can be written as 
\begin{equation}
\mathcal{S}(X)=X,\ \ \ \text{where }\mathcal{S}(X)\overset{\text{def}}{=} 
\mathrm{S}X\mathrm{S}^{-1}\ 
\end{equation}%
where $X$ is any acceptable operator, such as $H$ or $L$. For antiunitary $%
\mathrm{S}$, the induced map $\mathcal{S}$ is conjugate-linear in scalar
coefficients (e.g.\ $\mathcal{S}(iX)=\mathcal{S}(i)\mathcal{S}(X)$ with $%
\mathcal{S}(i)=-i$). The invariance of a superoperator $\mathcal{L}$
requires that 
\begin{equation}
\mathcal{S}\mathcal{L}=\mathcal{L}\mathcal{S}\text{ \ \ that is \ \ } 
\mathcal{S}\left( \mathcal{L}\left( X\right) \right) =\mathcal{L}\left( 
\mathcal{S}\left( X\right) \right)
\end{equation}%
for any $X$.

A commutator superoperator $\mathcal{C}_{H}(\rho )\overset{\text{def}}{=}%
[H,\rho ]$ is $S$-symmetric when $\mathcal{S}\mathcal{C}_{H}=\mathcal{C}_{H}%
\mathcal{S}$: 
\begin{equation}
\mathcal{S}\left( \mathcal{C}_{H}(\rho )\right) \equiv \left[ \mathcal{S}
\left( H\right) ,\mathcal{S}\left( \rho \right) \right]=\left[H,\mathcal{S}
\left( \rho \right) \right]\equiv \mathcal{C}_{H}\left( \mathcal{S}\left(
\rho \right) \right) \iff \mathcal{S}\left( H\right) =H
\end{equation}%
The same relation is valid for the double-commutator $\mathcal{C}%
_{L}^{2}(\rho )\overset{\text{def}}{=}\left[ L,[L,\rho ]\right]$: 
\begin{equation}
\mathcal{S}\left( \mathcal{C}_{L}^{2}(\rho )\right) \equiv \left[ \mathcal{S}
\left( L\right) ,\left[\mathcal{S}\left( L\right) ,\mathcal{S}\left( \rho
\right) \right]\right] =\left[ L,\left[L,\mathcal{S}\left( \rho \right) %
\right]\right] \equiv \mathcal{C}_{L}^{2}\left( \mathcal{S}\left( \rho
\right) \right) \iff \mathcal{S}\left( L\right) =L  \label{eq:CL2}
\end{equation}
(For pure dephasing, $\mathcal{S}(L)=\pm L$ leads to the same $\mathcal{C}%
^{2}_{L}$ because $L$ appears twice; the $\pm$ is omitted in (\ref{eq:CL2})
for notational simplicity.)

\subsection{Intrinsic (stochastic) dephasing under CP and CPT}

For the values used in (\ref{eq:Lam_stoc}) and in (\ref{eq:Lam_bath}), we
introduce the following CP-conjugate states and corresponding notations: 
\begin{equation}
\mathrm{U}_{\text{CP}}\left\vert k\right\rangle =\left\vert \bar{k}%
\right\rangle ,\ \ \ \mathrm{U}_{\text{CP}}\left\vert \bar{k}\right\rangle
=\left\vert k\right\rangle ,\ \ \ a_{\bar{j}}^{k}\overset{\text{def}}{=}%
a_{j}^{\bar{k}},\ b_{\bar{\beta}}^{k}\overset{\text{def}}{=}b_{\beta }^{\bar{%
k}}  \label{CPk}
\end{equation}%
where any phase factors associated with the CP transformation (which do not
affect the present analysis) are absorbed into $\mathrm{U}_{\text{CP}}$ and
the states to keep the notation simple. We assume a pure-dephasing setting
in which the coefficients $a_{j}^{k}$ and $b_{\beta }^{k}$ are real in the
chosen basis, and note that $\bar{\bar{k}}=k.$ Using these definitions one
finds 
\begin{equation}
\mathrm{U}_{\text{CP}}{L}_{j}\mathrm{U}_{\text{CP}}^{\dag }=\sum_{k}a_{j}^{k}%
\mathrm{U}_{\text{CP}}\left\vert k\right\rangle \left\langle k\right\vert 
\mathrm{U}_{\text{CP}}^{\dag }=\sum_{\bar{k}}a_{j}^{\bar{k}}\left\vert \bar{k%
}\right\rangle \left\langle \bar{k}\right\vert =L_{\bar{j}}  \label{LjCP}
\end{equation}%
The T-invariance of the operators $L_{j}$ in (\ref{symTHL}) also implies
that 
\begin{equation}
{\Theta }_{\text{CPT}}{L}_{j}{\Theta }_{\text{CPT}}^{-1}=\mathrm{U}_{\text{CP%
}}{\Theta }_{\text{T}}L_{j}{\Theta }_{\text{T}}^{-1}\mathrm{U}_{\text{CP}%
}^{\dag }=\pm \mathrm{U}_{\text{CP}}{L}_{j}\mathrm{U}_{\text{CP}}^{\dag
}=\pm {L}_{\bar{j}}  \label{LjCPT}
\end{equation}

The analysis of the stochastic (intrinsic) dephasing warrants a more
detailed consideration. Here symmetry covariance must be applied to the 
\textit{equation of motion} and must account for (i) a possible reversal of
the time parameter and (ii) the conjugate-linearity of antiunitary maps on
scalar factors such as $i$. For example, $S$-symmetry of the dephasing
Lindblad equation (\ref{eq:rho})--(\ref{eq:Lind}), assuming $\mathcal{S}%
\left( H\right) =H$ and defining $r=dt/d\tau =\pm 1$ (the minus sign
corresponds to time reversal $\tau =-t$), yields 
\begin{equation}
\mathcal{S}\left( i\hbar \frac{d\tilde{\rho}}{dt}\right) =\hbar \frac{%
\mathcal{S}(i)}{r}\frac{d\mathcal{S}(\tilde{\rho})}{d\tau }=\mathcal{S}%
\left( \mathcal{C}_{H}\left( \tilde{\rho}\right) -\mathcal{D}\tilde{\rho}%
\right) =\left[ H,\mathcal{S}(\tilde{\rho})\right] -\mathcal{S}\left( 
\mathcal{D}\tilde{\rho}\right)
\end{equation}%
implying that symmetry of the equation requires 
\begin{equation}
\mathcal{S}\left( \mathcal{D}\tilde{\rho}\right) =\mathcal{D}\left( \mathcal{%
S}\tilde{\rho}\right)
\end{equation}%
since $\mathcal{S}(i)/r=+1$ for both CP and CPT transformations.

The CP and CPT symmetries impose different requirements for the dephasing
operator $\mathcal{D}(\tilde{\rho})$ specified in (\ref{eq:Lam_stoc}): 
\begin{equation}
\text{CP: \ \ \ }\mathcal{S}_{\text{CP}}\left( \mathcal{D}(\tilde{\rho}%
)\right) =\mathcal{D}\left( \mathcal{S}_{\text{CP}}\left( \tilde{\rho}%
\right) \right) \Longrightarrow \gamma _{j}=\gamma _{\bar{j}}\text{ for all }%
j  \label{CPgam}
\end{equation}%
\begin{equation}
\text{CPT: \ \ \ }\mathcal{S}_{\text{CPT}}\left( \mathcal{D}(\tilde{\rho}%
)\right) =\mathcal{D}\left( \mathcal{S}_{\text{CPT}}\left( \tilde{\rho}%
\right) \right) \Longrightarrow \gamma _{j}=-\gamma _{\bar{j}}\text{ for all 
}j  \label{CPTgam}
\end{equation}%
The last transformation takes into account (\ref{symTHL}), (\ref{eq:CL2}), (%
\ref{LjCP}) and (\ref{LjCPT}) and the antiunitary character of time
reversal: 
\begin{equation}
{\Theta }_{\text{CPT}}{(i)}{\Theta }_{\text{CPT}}^{-1}=\Theta _{\text{T}}{(i)%
}{\Theta }_{\text{T}}^{-1}=\mathrm{K}{(i)}\mathrm{K}=(i)^{\ast }=(-i)
\end{equation}

\subsection{Environmental dephasing and effective invariance}

As previously noted in (\ref{symTHL}), the Lindbladian bath-interaction
operators ${L}_{\beta }^{B}$ are assumed to be T-invariant and, therefore, 
\begin{equation}
\mathrm{U}_{\text{CP}}{L}_{\beta }^{B}\mathrm{U}_{\text{CP}}^{\dag }=\sum_{%
\bar{k}}b_{\beta }^{\bar{k}}\left\vert \bar{k}\right\rangle \left\langle 
\bar{k}\right\vert =L_{\bar{\beta}}^{B},\qquad {\Theta }_{\text{CPT}}{L}%
_{\beta }^{B}{\Theta }_{\text{CPT}}^{-1}=\pm {L}_{\bar{\beta}}^{B},
\end{equation}%
in line with equations (\ref{CPk})--(\ref{LjCPT}). Since we assume CP-, CPT-
and T-invariance of all Hamiltonians used in (\ref{eq:HSB_decomposition}),
the \textit{microscopic} dynamics of the system--bath composite is invariant
under these transformations and represents exactly the same problem as the
one solved in Section~\ref{subsec:bath_dephasing}; that is environmental
interactions are deemed to propagate through charge-neutral channels.
Considering that the bath-induced dephasing rates $\gamma _{\beta }^{B}$ are
determined solely by the bath spectra in (\ref{eq:gamma_eta}), the
invariance of the full composite implies that conjugate channels probe the
same bath noise strength, hence 
\begin{equation}
\gamma _{\beta }^{B}=\gamma _{\bar{\beta}}^{B}.  \label{Bgam}
\end{equation}

Importantly, the \textit{time-directionality} of the reduced
Lindblad/Redfield equation does not originate from any CP, CPT, or T
violation of the Hamiltonians: it enters through additional closure
assumptions (Born decorrelation and the Markov/short-correlation limit),
which are conceptually analogous to Boltzmann's molecular chaos and reflect
the empirical dominance of decoherence over recoherence in our Universe.
These assumptions select the forward-time semigroup form for the \textit{%
reduced} dynamics. This leaves the microscopic CP-covariant relation (\ref%
{Bgam}) intact---compare this relation with (\ref{CPgam}) and (\ref{CPTgam}%
)---but, because of temporal directionality of the mechanism under
consideration, the reduced description tends not to remain T-invariant and
consequently need not display naive CPT covariance. Equation~(\ref{Bgam}) is
valid irrespective of whether (\ref{CPgam}) or (\ref{CPTgam}) is adopted.

We note that the systems considered here are open: the system interacts with
the bath, and the bath interacts with the wider Universe. The CP and CPT
transformations of an open system are therefore incomplete: they involve
charge conjugation in the system, $\left\vert k\right\rangle \leftrightarrow
\left\vert \bar{k}\right\rangle $, but not in the bath or in the rest of the
Universe. A complete C transformation would require replacing matter by
antimatter in the bath and, by extension, in the wider Universe, which is
not physically realisable. Hence an incomplete symmetry may, but need not,
be violated without undermining the corresponding complete symmetry. Even if
complete CPT invariance is expected to remain conceptually valid, incomplete
CPT transformations in open systems can readily be violated by external
influences: interventions from the wider environment may break the symmetry
of the effective open-system dynamics, producing apparent CPT asymmetries
even when the underlying microscopic Hamiltonian of the system and the
system's intrinsic properties are CPT-invariant \cite{K-PhysA}.

\bigskip

The overall dephasing equation, which includes both intrinsic and
environmental contributions and is written in the energy eigenbasis, is 
\begin{equation}
i\hbar \frac{d\tilde{\rho}_{ki}}{dt} = \left( E_{k}-E_{i}\right) \tilde{\rho}%
_{ki} -i\left( \Gamma _{ki}+\Gamma _{ki}^{B}\right) \tilde{\rho}_{ki},
\end{equation}
where 
\begin{equation}
\Gamma _{ki} = \frac{1}{2\hbar }\sum_{j}\gamma _{j} \left(
a_{j}^{k}-a_{j}^{i}\right) ^{2}, \qquad \Gamma _{ki}^{B} = \frac{1}{2\hbar }%
\sum_{\beta }\gamma _{\beta }^{B} \left( b_{\beta }^{k}-b_{\beta
}^{i}\right) ^{2},
\end{equation}
according to (\ref{eq:Lam_stoc}) and (\ref{eq:Lam_bath}). The bath dephasing
term is invariant under conjugation of both indices: 
\begin{equation}
\Gamma _{ki}^{B}=\Gamma _{\bar{k}\bar{\imath}}^{B}.
\end{equation}
Similar relations apply to the stochastic dephasing term under CP
invariance, whereas under CPT invariance the sign is reversed: 
\begin{eqnarray}
\text{CP:}\ \ \ &&\Gamma _{ki}=\Gamma _{\bar{k}\bar{\imath}},  \label{CPGki}
\\
\text{CPT:}\ \ &&\Gamma _{ki}=-\Gamma _{\bar{k}\bar{\imath}}.  \label{CPTGki}
\end{eqnarray}

\subsection{Two-particle channels and symmetry consequences}

As a more specific example, we consider a two-particle initial state $%
\left\vert qr\right\rangle _{1}=\left\vert q\right\rangle _{1}\otimes
\left\vert r\right\rangle _{1}$ that is transformed into the final state $%
\left\vert qr\right\rangle _{2}=\left\vert q\right\rangle _{2}\otimes
\left\vert r\right\rangle _{2}$. With notation $\Gamma(k,i)\overset{\text{%
{\tiny def}}}{=}\Gamma _{ki}$, the CP-covariant relations take the form 
\begin{equation}
\text{CP:}\ \ \Gamma \left( \left\vert q_{1}r_{1}\right\rangle ,\left\vert
q_{2}r_{2}\right\rangle \right) =\Gamma \left( \left\vert \bar{q}_{1}\bar{r}%
_{1}\right\rangle ,\left\vert \bar{q}_{2}\bar{r}_{2}\right\rangle \right) .
\label{CPG}
\end{equation}

The CPT case requires more detailed consideration, with matter states
denoted by $\left\vert p\right\rangle $, antimatter states by $\left\vert 
\bar{p}\right\rangle $, and neutral states by $\left\vert o\right\rangle $: 
\begin{equation}
\mathrm{U}_{\text{CP}}\left\vert p\right\rangle =\left\vert \bar{p}%
\right\rangle ,\qquad \mathrm{U}_{\text{CP}}\left\vert \bar{p}\right\rangle
=\left\vert p\right\rangle ,\qquad \mathrm{U}_{\text{CP}}\left\vert
o\right\rangle =\left\vert \bar{o}\right\rangle =\left\vert o\right\rangle .
\end{equation}%
We take into account that the transformation $\left\vert \cdots
\right\rangle _{1}\rightarrow \left\vert \cdots \right\rangle _{2}$ must
preserve charge throughout the transition; that is, a matter state cannot be
converted into an antimatter state. It is assumed that only conventional
non-negative $\gamma _{j}$ can be associated with common (matter) states.
The symmetry relation (\ref{CPTGki}) is then written as 
\begin{equation}
\text{CPT:}\ \ \Gamma \left( \left\vert pp\right\rangle _{1},\left\vert
pp\right\rangle _{2}\right) =-\Gamma \left( \left\vert \bar{p}\bar{p}%
\right\rangle _{1},\left\vert \bar{p}\bar{p}\right\rangle _{2}\right) \geq 0,
\label{CPTG1}
\end{equation}%
\begin{equation}
\text{CPT:}\ \ \Gamma \left( \left\vert p\bar{p}\right\rangle
_{1},\left\vert p\bar{p}\right\rangle _{2}\right) =-\Gamma \left( \left\vert 
\bar{p}p\right\rangle _{1},\left\vert \bar{p}p\right\rangle _{2}\right) 
\overset{\text{sym}}{=}0,  \label{CPTG2}
\end{equation}%
\begin{equation}
\text{CPT:}\ \ \Gamma \left( \left\vert po\right\rangle _{1},\left\vert
po\right\rangle _{2}\right) =-\Gamma \left( \left\vert \bar{p}o\right\rangle
_{1},\left\vert \bar{p}o\right\rangle _{2}\right) \geq 0,  \label{CPTG3}
\end{equation}%
\begin{equation}
\text{CPT:}\ \ \Gamma \left( \left\vert oo\right\rangle _{1},\left\vert
oo\right\rangle _{2}\right) =-\Gamma \left( \left\vert \bar{o}\bar{o}%
\right\rangle _{1},\left\vert \bar{o}\bar{o}\right\rangle _{2}\right) =0.
\label{CPTG4}
\end{equation}%
The label \textquotedblleft sym\textquotedblright\ in equation~(\ref{CPTG2})
indicates that the physical problem is assumed to be symmetric under
interchange of the two particles; under this assumption, the corresponding
expression vanishes.

\bigskip

We note that CPT-invariant decoherence tends to produce the most pronounced
contrast between matter--neutral and antimatter--neutral dephasing channels
due to intrinsic decohering neutrality of the neutral states \cite%
{Klimenko2017KineticsCPT}. Under CP and CPT symmetries, we have different
predictions for the total dephasing between the antistate and the neutral
state: 
\begin{eqnarray}
\text{CP}\text{: } &&\Gamma _{\bar{p}o}^{\text{tot}}=\Gamma _{po}^{B}+\Gamma
_{po}=\Gamma _{po}^{\text{tot}}  \label{GtotCP} \\
\text{CPT}\text{: } &&\Gamma _{\bar{p}o}^{\text{tot}}=\Gamma
_{po}^{B}-\Gamma _{po}<\Gamma _{po}^{\text{tot}}  \label{GtotCPT}
\end{eqnarray}%
where we denote $\Gamma ^{\text{(\#)}}\left( \left\vert qo\right\rangle
_{1},\left\vert qo\right\rangle _{2}\right) $ by $\Gamma _{qo}^{\text{(\#)}}$
for all superscripts $\#$ and $q=p$ or $\bar{p}$.\ \textsl{Note that the
relative decoherence factor }$\phi =\Gamma _{\bar{p}o}^{\text{tot}}/
\Gamma_{po}^{\text{tot}}$\textsl{\ defined after (\ref{eq:sd_Phi_asymptote}) is }$%
\phi =1$\textsl{\ under CP and\ }$\phi <1$\textsl{\ under CPT conditions}.
Allowing some of the coefficients $\gamma _{j}$ to take negative values is,
in many cases, only an intermediate step: its net effect is to reduce the
positive total dephasing rate $\Gamma ^{\mathrm{tot}}$, rather than to
generate sustained recoherence.

\section{Diffractive dissociations and decoherence}

This section considers possible decoherence--interference effects in \textit{%
single-} and \textit{double-diffractive} dissociation (SD and DD) \cite%
{GoodWalker1960,PDGSoftQCD2021,HEPDiffraction2002} in high-energy hadronic
collisions \cite{PDG2024_RPP}. Diffractive collisions are characterised by
small momentum transfer and the dissociation of one or both incoming hadrons
into multiparticle systems separated by a large rapidity gap (SD: $p+X$; DD: 
$X_{1}+X_{2}$) \cite{PDGSoftQCD2021,DeWolf2002}. These events necessarily
involve loss of coherence between the separated hadronised systems and an
associated increase of entropy, but they do not produce a single strongly
interacting medium: the rapidity gap signals the absence of substantial
thermalisation. In this sense, the primary irreversible element relevant
here is \textit{decoherence} between the outgoing systems rather than
equilibration. \textsl{The very presence of decoherence points to
non-unitary effects naturally characterised by a dephasing Lindblad equation 
\cite{Lindblad1976,Breuer2007book,LindbladIntro2020}. }

Soft diffractive observables are commonly treated in \textit{Regge theory}
in terms of colourless vacuum exchange, the \textit{Pomeron} ($\mathbb{P}$),
supplemented at lower energies by subleading crossing-odd trajectories (%
\textit{Reggeons}) \cite{PDGSoftQCD2021,DeWolf2002}. As energy increases,
Reggeon contributions fall while $\mathbb{P}$ exchange dominates, so
diffractive processes become increasingly insensitive to the replacement of
protons by antiprotons $p\rightarrow \bar{p}$; this is consistent with the
general trend toward asymptotic similarity of $pp$ and $p\bar{p}$ reactions 
\cite{Pomeranchuk1958,DonnachieLandshoff1992}. Regge descriptions may be
viewed as an effective, confinement-insensitive parameterisation of soft
exchange dynamics (i.e. enabling long-range interactions between hadrons
without an explicit treatment of confined coloured degrees of freedom),
although their precise relation to Quantum Chromodynamics (QCD) is a complex
matter.

The main goal of the rest of this work is to examine published data on
proton--proton ($pp$) and proton--antiproton ($p\bar{p}$) collisions,
focusing on diffractive cross-sections, to assess whether the available
evidence favours CP- or CPT-invariant forms of the generalised dephasing
Lindblad equation.

\subsection{Interaction diagrams and interference with decoherence}

According to the Good--Walker formalism \cite{GoodWalker1960,PDGSoftQCD2021}%
, diffractive scattering may be viewed as involving two principal
ingredients: (i) a unitary interaction described by the transition matrix,
and (ii) an effective loss of coherence associated with the dissociation of
at least one incoming hadron into diffractive final states. This loss of
coherence is essential: if the incoming hadron were a single diffractive
eigenstate (no fluctuations among eigencomponents), only elastic scattering
would occur; diffractive dissociation appears when different eigencomponents
scatter with different amplitudes and decohere.

Diffractive cross-sections are conventionally evaluated using cut diagrams
based on Mueller's optical theorem \cite{HEPDiffraction2002}. Such diagrams,
however, are merely graphical representations of unitary identities and
therefore neither describe nor permit intrinsically non-unitary evolution.
The cut diagrams point to central significance of triple-Pomeron vertex.
Here we aim to incorporate decoherence into the conventional framework and
assess how it modifies the standard expectations. \textsl{The Good--Walker
decoherence and triple-Pomeron are often seen as different views on the same
phenomenon \cite{Gustafson2012}}.

\begin{figure}[h]
\begin{center}
\includegraphics[width=14cm,page=1,trim=0.5cm 5cm 2cm 4.5cm, clip ]{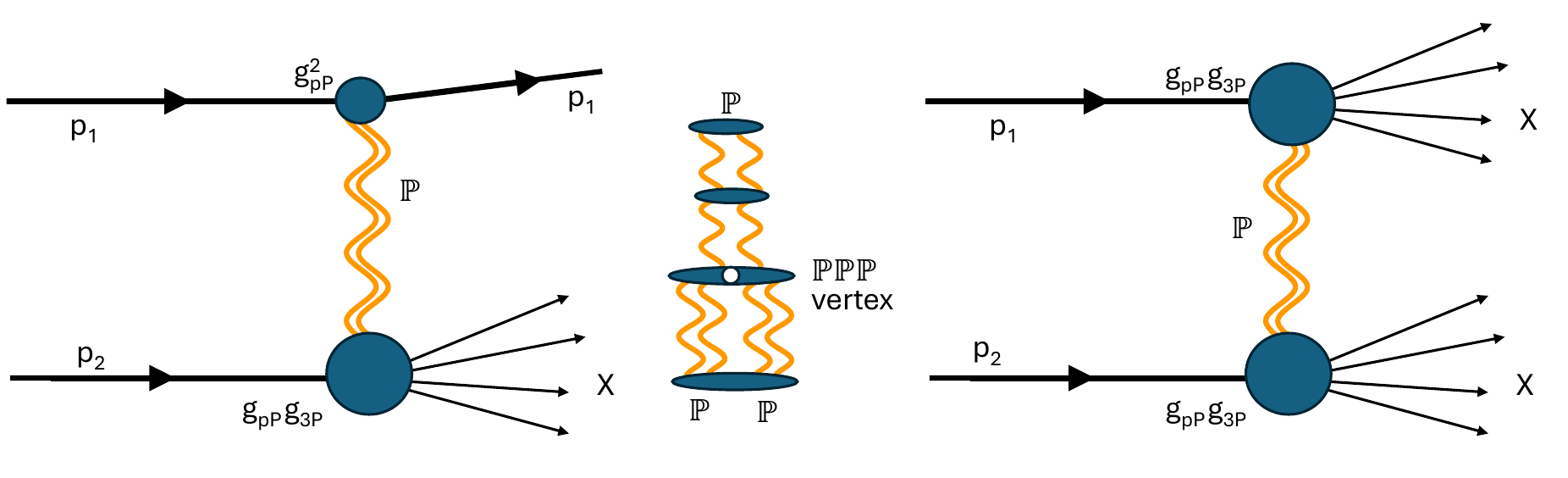}
\caption{Interaction diagrams for proton (antiproton) single diffraction (SD, left) and double diffraction (DD, right) via Pomeron $\mathbb{P}$ exchange, often sketched as two-gluon colour-singlet. The central inset  sketches the QCD-motivated ladder picture of the Pomeron, and its split at the triple-$\mathbb{P}$ vertex.}
\label{fig1}
\end{center}
\end{figure}

Figure \ref{fig1} shows schematic uncut (amplitude-level) topologies for
single diffraction (SD, left) and double diffraction (DD, right) mediated by
Pomeron exchange. The central inset sketches the commonly used QCD-motivated
picture of the Pomeron as a gluonic ladder and indicates how ladder
splitting generates the familiar triple-Pomeron ($\mathbb{P}\mathbb{P}%
\mathbb{P}$) configuration \cite{HEPDiffraction2002}. In purely elastic
scattering, the interaction is represented by a single Pomeron exchange
between intact hadrons. In high-mass SD, the dominant contribution is
conventionally described by a triple-Pomeron topology, in which the
exchanged Pomeron effectively splits into two, with the $\mathbb{P}\mathbb{P}%
\mathbb{P}$ vertex attached to the dissociating side. Higher-order
topologies containing multiple $\mathbb{P}\mathbb{P}\mathbb{P}$ vertices
may, in principle, also contribute, but the leading triple-Pomeron term is
typically sufficient for SD, where the opposite hadron remains intact. In
high-mass DD, dissociation occurs on both sides and the corresponding
triple-Regge topology is commonly represented by two $\mathbb{P}\mathbb{P}%
\mathbb{P}$ vertices.

The interplay between hadronic degrees of freedom and the Pomeron ladder
lies at the heart of diffractive dissociation dynamics. The SD and DD
cross-sections corresponding to Figure~\ref{fig1} can be written in the
conventional triple-Regge form as products of effective couplings and Regge
factors, as obtained from Mueller's optical theorem \cite%
{Collins1977_ReggeTheory,PDGSoftQCD2021}.

\begin{align}
\sigma _{_{\mathrm{SD}}}& \propto \,g_{\mathbb{P}p_{1}}^{2}\,g_{\mathbb{P}%
p_{2}}^{{}}\,g_{3\mathbb{P}(p_{2})}^{{}}\,F\left( s,t,M_{X}^{2}\right) ,
\label{sigSD} \\
\sigma _{_{\mathrm{DD}}}& \propto \,g_{\mathbb{P}p_{1}}^{{}}\,g_{3\mathbb{P}%
(p_{1})}^{\,}\,g_{\mathbb{P}p_{2}}^{{}}\,g_{3\mathbb{P}(p_{2})}^{\,}\,F%
\left( s,t,M_{X_{1}}^{2},M_{X_{2}}^{2}\right) ,  \label{sigDD}
\end{align}%
where $g_{\mathbb{P}p}^{{}}$ and $g_{3\mathbb{P}}^{{}}$ are effective
couplings, defined and discussed below; $F(\ldots )$ denotes the remaining
kinematic dependence (including the appropriate integrations over and/or
dependence on central energy squared $s$, four-momentum exchange $t$ and
invariant mass $M_{X}$ of the dissociated system $X$). In SD, particle $%
p_{1} $\ remains intact and particle $p_{2}$ dissociates. The
transverse-momentum\ dependence of $g_{\mathbb{P}p}^{{}}$ on the
intact-proton side, or the effective transverse-momentum independence of $g_{%
\mathbb{P}p}^{{}}$ on the dissociation side, is not germane to the present
discussion and is therefore not indicated.\ The argument $(p)$ in
triple-Pomeron coupling indicates that $g_{3\mathbb{P}(p)}^{{}}$ is
associated with dissociating particle $p$ and is affected by this particle.
The coupling associated with $p_{2}$, i.e. $G_{\mathbb{PPP}}^{p_{2}}=g_{%
\mathbb{P}p_{2}}^{{}}\,g_{3\mathbb{P}(p_{2})}^{{}}$ is often interpreted as
the overall coupling on the dissociating side.

From a QCD perspective, the Pomeron is an effective colour-neutral exchange
dominated by gluonic degrees of freedom and carrying vacuum quantum numbers.
It is not a stable particle and, in appropriate limits, may be represented
as a ladder-like gluonic system whose evolution includes branching and
recombination. The partonic substructure of hadrons, normally hidden by QCD
confinement, can manifest itself in diffractive dissociation through such
Pomeron systems, so that the ladder evolution is effectively stochastic at
the partonic level.\textsl{\ In this sense, QCD-based diagrams may be viewed
as representing effective interaction topologies and physical couplings.} 
\textsl{In principle, a single Pomeron can couple to two or more partons.
However, a triple-Pomeron splitting, }$\mathbb{P}\rightarrow \mathbb{P}+%
\mathbb{P}$\textsl{, allows the lower part of the interaction to be
represented by two more independent ladders, each capable of evolving and
coupling separately \cite{Bartels2017}. In this sense, the triple-Pomeron
topology provides the basic, leading-order configuration associated with
independent evolution and loss of coherence on the dissociating side. From
this perspective, a two-Pomeron configuration is more naturally associated
with the dissociating proton (although a single Pomeron may still couple to
more than one parton while retaning coherence of these interactions). }

\subsection{The decoherence factors}

The two versions of the dephasing operator in the Lindblad equation lead to
distinct experimental expectations. In the CP-invariant version (\ref{GtotCP}%
), matter and antimatter contribute identically to decoherence, whereas in
the CPT-invariant version (\ref{GtotCPT}) their contributions differ; in
principle, this difference is experimentally accessible.

According to our CPT-invariant analysis (\ref{CPTG4}), self-conjugate
bosonic excitations---objects that coincide with their antiparticles, such
as gluonic exchanges or radiation---must be decoherence-neutral (consistent
with earlier thermodynamic arguments \cite{Klimenko2017KineticsCPT}). That
is, they may behave stochastically and can mediate or amplify decoherence by
opening additional diffractive channels, but they cannot by themselves
discriminate between forward-time (decohering) and reverse-time (recohering)
stochastic evolution; any such bias must originate from the
matter--antimatter sector.

In the superoperator formulation of the perturbative Dyson expansion,
adapted to dephasing dynamics in Section~\ref{subsec:sd_twochannel_lindblad}%
, the coupling $g$ is multiplied by a decoherence factor $\Phi $ that
depends on the total dephasing strength $\Gamma ^{\text{tot}}=\Gamma ^{B}\pm
\Gamma $, according to (\ref{eq:sd_Phi_asymptote}), (\ref{GtotCP}) and (\ref%
{GtotCPT}). In this convention, the \textquotedblleft $+$\textquotedblright\
sign corresponds to the CP-invariant treatment (and to particles in the
CPT-invariant interpretation), whereas the \textquotedblleft $-$%
\textquotedblright\ sign corresponds to the CPT-invariant contribution
associated with antiparticles. We therefore expect interactions involving
antiparticles to incur an additional decoherence penalty, reflecting the
recohering tendency attributed to antimatter in CPT-invariant dephasing.

A convenient measure of this penalty is relative decoherence factor%
\begin{equation}
\phi =\frac{\Phi _{\bar{p}\mathbb{P}}}{\Phi _{p\mathbb{P}}}=\frac{\Gamma _{p%
\mathbb{P}}^{B}-\Gamma _{p\mathbb{P}}}{\Gamma _{p\mathbb{P}}^{B}+\Gamma _{p%
\mathbb{P}}}\leq 1,  \label{fiGG}
\end{equation}%
which quantifies the attenuation of effective couplings for antimatter,
relative to matter, induced by decoherence under CPT invariance \textsl{as
determined by (\ref{eq:sd_Phi_asymptote}) and (\ref{GtotCPT})}. We take this
factor to enter multiplicatively for each relevant Pomeron coupling to
matter or antimatter. Accordingly, in line with \textsl{(\ref%
{eq:sd_Phi_asymptote}), we write}%
\begin{equation}
g_{\mathbb{P}\bar{p}}=\phi \,g_{\mathbb{P}p},\qquad g_{3\mathbb{P}(\bar{p}%
)}=\phi ^{3}\,g_{3\mathbb{P}(p)},  \label{eqfi}
\end{equation}%
where the first relation modifies the direct $\mathbb{P}$--antiparticle
coupling and the second modifies the triple-$\mathbb{P}$ coupling anchored
to a dissociating $\bar{p}$. \ \textsl{Taking into account the possibility
of \ }$\mathbb{P}\rightarrow 2\mathbb{P}$\textsl{\ splitting, and noting
that a two-ladder configuration entails more independent evolution \cite%
{Bartels2017}, we take it in the present context to correspond to stronger
dephasing effects. As a single triple-}$\mathbb{P}$ \textsl{vertex is the
basic configuration for SD, the effective number of Pomeron--parton
interactions is expected to be approximately twice as large on the
dissociating side.} We accordingly assume that, on each side, the number of $%
\phi $-factor multipliers is proportional to the effective number of
Pomeron--parton interactions, and hence is also twice as large on the
dissociating side. For $\phi =1$, these expressions reduce to the proton
case and may also be used for both $p$ and $\bar{p}$ in the CP-invariant
limit.

\subsection{Effect on diffractive dissociation reactions}

Single diffractive (SD) events can be divided into the following four
groups: \ A) proton dissociation in $pp$ collisions, B) proton dissociation
in $p\bar{p}$ collisions, C) antiproton dissociation in $p\bar{p}$
collisions, and D) antiproton dissociation in $\bar{p}\bar{p}$ collisions,
represented by the corresponding reactions 
\begin{gather}
\text{A) }p+p\rightarrow p+X,\ \ \ \text{B) }\bar{p}+p\rightarrow \bar{p}+X,
\notag \\
\text{C)\ }p+\bar{p}\rightarrow p+\bar{X},\ \ \text{\ D)\ }\bar{p}+\bar{p}%
\rightarrow \bar{p}+\bar{X}  \label{rc1}
\end{gather}%
with the corresponding cross-sections $\sigma _{{_{\mathrm{SD}}}}$ denoted
by $\sigma _{\text{{\tiny A}}},$ $\sigma _{\text{{\tiny B}}}$, $\sigma _{%
\text{{\tiny C}}},$ and $\sigma _{\text{{\tiny D}}}$. Equations (\ref{sigSD}%
), (\ref{fiGG}) and (\ref{eqfi}) indicate that 
\begin{equation}
\frac{\sigma _{\text{{\tiny B}}}}{\sigma _{\text{{\tiny A}}}}=\phi ^{2},\ \
\ \frac{\sigma _{\text{{\tiny C}}}}{\sigma _{\text{{\tiny A}}}}=\phi ^{4},\
\ \ \frac{\sigma _{\text{{\tiny D}}}}{\sigma _{\text{{\tiny A}}}}=\phi ^{6}
\label{sigSD3}
\end{equation}%
due to the expected progressive partial suppression of the
decoherence-induced coupling enhancement associated with antiprotons---a
factor $\phi $ is multiplicatively included for each effective Pomeron
interaction on the antiproton side.

Similarly, double diffractive (DD) events are divided into the following
three groups associated with\ E) $pp$ collisions, F) $p\bar{p}$ collisions,
and G) $\bar{p}\bar{p}$ collisions as represented by the corresponding
reactions 
\begin{equation}
\text{E) }p+p\rightarrow X+X,\ \ \ \text{F)\ }p+\bar{p}\rightarrow X+\bar{X}%
,\ \ \text{\ G)\ }\bar{p}+\bar{p}\rightarrow \bar{X}+\bar{X}  \label{rc2}
\end{equation}%
with the corresponding cross-sections $\sigma _{{_{\mathrm{DD}}}}$ denoted
by $\sigma _{\text{{\tiny E}}},$ $\sigma _{\text{{\tiny F}}},$ and $\sigma _{%
\text{{\tiny G}}}$. Equations (\ref{sigDD}), (\ref{fiGG}) and (\ref{eqfi})
indicate that 
\begin{equation}
\frac{\sigma _{\text{{\tiny F}}}}{\sigma _{\text{{\tiny E}}}}=\phi ^{4},\ \
\ \frac{\sigma _{\text{{\tiny G}}}}{\sigma _{\text{{\tiny E}}}}=\phi ^{8}
\label{sigDD2}
\end{equation}%
reflecting the expected progressive partial suppression of two-particle
decoherence by interactions of antiparticles and Pomerons.

Note that the expected value of $\phi <1$ is specifically linked to
non-unitary effects of CPT-invariant priming of decoherence linked to
stochastic realism. The existing unitary theories and conventional QCD
predict $\phi =1$. The same expectation $\phi =1$ is associated with
CP-invariant interpretation of the Lindblad equation: while decoherence may
influence the effective coupling, these influences must be the same for
particles and antiparticles and would be hard to detect.

\section{Experimental cross-sections for diffractive collisions}

We now turn to examining the experimental values of the relative decoherence
factor $\phi $. Although dedicated measurements in $\bar{p}+\bar{p}$
collisions have not been performed, published results are available for both
SD and DD in $\bar{p}+p$ and $p+p$ collisions, albeit very sparse for DD and
sufficient---but still limited---for SD.

Diffractive dissociations are a class of inelastic interactions
characterised by the exchange of a low-momentum transfer resulting in at
least one of the colliding particles dissociating into a low-mass system of
secondary particles while maintaining a very small angular deflection
relative to the original beam direction. The smallness of these deflections
makes capturing such events challenging, introducing substantial uncertainty
into measurements. Roman pot detectors are designed to move these detectors
as close as possible to the beam \cite{Battiston1985_UA4RomanPots}. Still, a
complete accounting of diffractive dissociation events is among the most
challenging aspects of inelastic-collision measurements, particularly at
higher energies where the characteristic deflection angles become even
smaller. In many collider experiments, SD events are excluded from the
headline inelastic samples in order to avoid the associated uncertainties
(hence the frequent use of non-single-diffractive, NSD, event classes).
Measurements of diffractive cross sections typically require an
extrapolation into the region of very small deflection angles---and
therefore low diffractive dissociation masses $M_{X}$---where the direct
experimental acceptance is limited. As experiments move to progressively
higher collision energies this is beneficial in many respects; however,
determining the full SD cross section becomes increasingly challenging while
resorting to simulations or reporting only partial (fiducial) values may
become unavoidable.

In the context of the present work, a range of extrapolation strategies may
be employed but with an important caveat. Baseline models typically (often
implicitly) presume $\phi =1$ and therefore cannot provide an independent
test of $\phi \neq 1$. It is therefore essential that modelling is not used
as a stand-alone tool or as a direct substitute for missing data. Models may
only be used to help guide the shape of an interpolation or extrapolation,
but the experimental data must continue to determine the overall magnitude,
rather than being replaced by model predictions in poorly constrained
regions; otherwise, the inference can become model-based or model-dominated.
Likewise, parameters inferred from other measurements---especially those
corresponding to a different reaction category---should not be imported or
conflated: only experiments for which the selected reaction channel can be
identified unambiguously are suitable. Nevertheless, to enable consistent
comparisons across energies, the measured SD cross sections must be extended
to the full $M_{X}$ range, down to the kinematic limit $M_{X}\sim m_{p}$, as
is commonly done in historical measurements.

Given the inherent uncertainties in diffractive dissociation measurements,
the most reliable approach would be a direct comparison of proton and
antiproton dissociation measured within the same experiment using an
identical methodology. Unfortunately, such side-separated results are very
scarce---almost non-existent---in the published literature. In many cases,
the reported total diffractive cross-sections include dissociation on both
sides of the interaction; however, a careful reading of the experimental
procedures indicates that the measurements were often performed on one side
only, with the quoted totals obtained by doubling the measured one-sided
values. As a result, a proton--antiproton comparison is presently feasible
only by combining results from different experiments. This
inter-experimental approach can introduce additional systematic
uncertainties associated with differences in detector acceptance, trigger
selection, and analysis methodology. Increasing the number of data points
and prioritising the most reliable measurements can help to mitigate these
effects.

\subsection{Compiled data and experimental notes}

\begin{figure}[h]
\begin{center}
\includegraphics[width=14cm,page=1,trim=1.5cm 4cm 1.5cm 8cm, clip ]{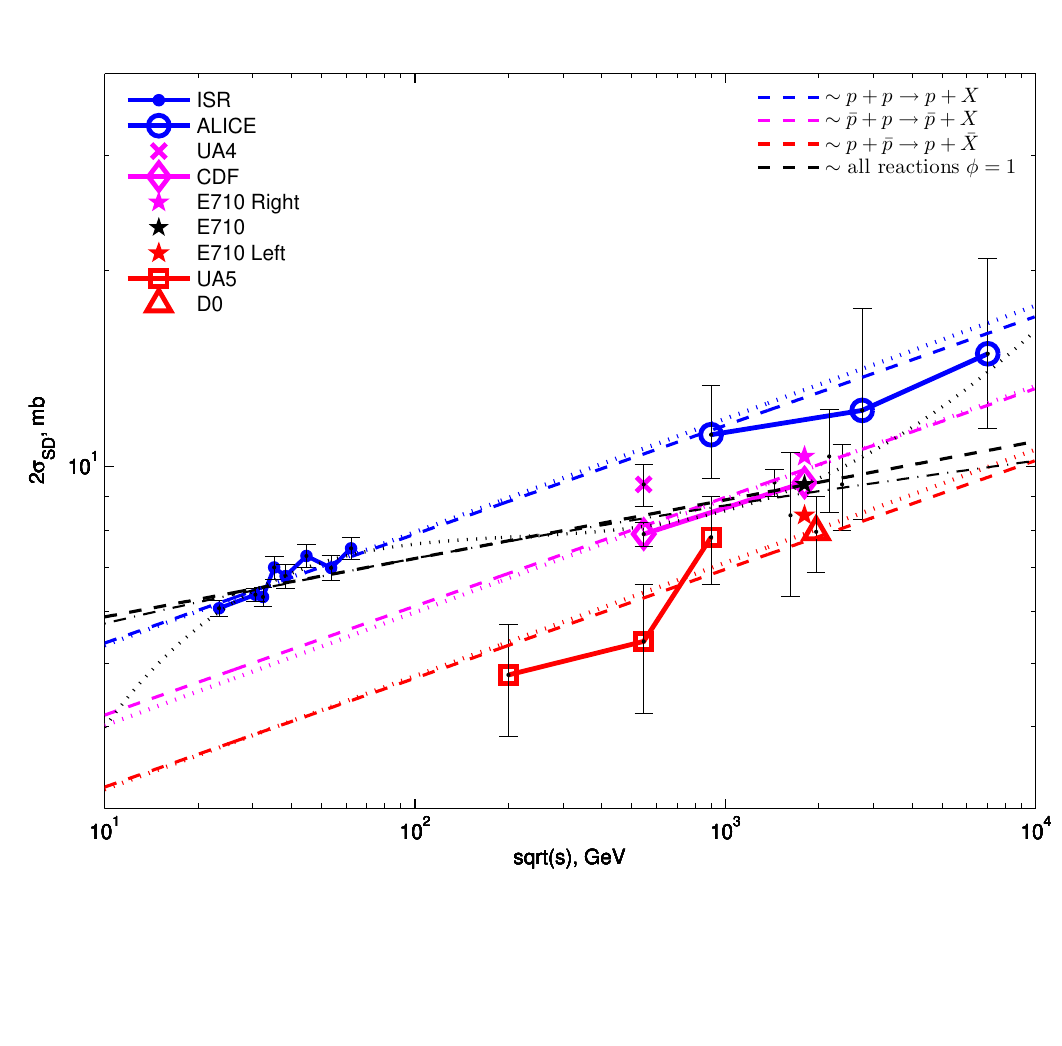}
\caption{SD cross-section $2\sigma _{\mathrm{SD}}$ vs $\sqrt{(s)}$. Experimental data are from ISR,
UA4, UA5, CDF, E710, D0 and ALICE collaborations. Approximations: \  -- \ -- \ --
SDF1 (colour) or SDC1 (black),\ \  $\cdot $ $\cdot $ $\cdot $ $\cdot $ $%
\cdot $ $\cdot \ $ SDF2 (colour) or SDC4 (black), \ $\cdot $ -- $\cdot $ -- $%
\cdot $ -- SDCln. 
}
\label{fig2}
\end{center}
\end{figure}

Figure \ref{fig2} shows the SD cross-sections for a broad range of collision
energies. Note that Figure \ref{fig2} (and the rest of this work) shows $%
2\sigma _{_{\mathrm{SD}}}$ in line with the conventional doubling of the
cross-section, assuming the same cross sections for SD on the left-hand and
right-hand sides of the experiment. While we obviously avoid making this
assumption and analyse one-sided characteristics, we still follow the
doubling convention and report $2\sigma _{_{\mathrm{SD}}}$ for ease of
comparison with the published literature. When different errors (e.g.
statistical, systematic) are reported (statistical errors are listed first)
or evaluated separately, the error bars show the RMS sum of all errors (i.e.
assuming that the errors are independent).

The \textbf{ISR} experiments were conducted at lower energies and this makes
SD measurements easier despite using older equipment---one can see that the
error bars reported in Figure \ref{fig2} for ISR experiments are relatively
narrow. These experiments were of type~A, with an exception of a preliminary
p\={p} run that was focused on central rapidity and is not suitable for SD
analysis. The reported values of $2\sigma _{_{\mathrm{SD}}}$ range between $%
6 $ and $7.5$~mb for energies $30\leq \sqrt{s}\leq 65$~GeV \cite%
{CHLM1976,Armitage1982}.

\textbf{UA5} \cite{UA5Ansorge1986} reports $2\sigma _{_{\mathrm{SD}}}=7.8\pm
0.5\pm 1.1$~mb at 900~GeV and $2\sigma _{_{\mathrm{SD}}}=4.8\pm 0.5\pm 0.8$%
~mb at 200~GeV, while identifying acceptance tuned to type~C events. The UA5
value $2\sigma _{_{\mathrm{SD}}}=5.4\pm 1.1$ at 546~GeV is mentioned as less
certain, but is not explicitly reported in UA5~\cite{UA5Ansorge1986} and is
commonly taken from secondary compilations (see, e.g., discussion in \cite%
{Abramowicz2000,Poghosyan2010}); it therefore appears less reliable than the
two values reported for 200 and 900~GeV. \textbf{UA4} \cite{UA4Bernard1987}
reports $2\sigma _{_{\mathrm{SD}}}=9.4\pm 0.7$~mb at 546~GeV while
identifying reaction~B as the target.

\textbf{CDF} \cite{CDF_Abe1994_SD} quotes $2\sigma _{_{\mathrm{SD}}}=7.89\pm
0.33$~mb at 546~GeV and $2\sigma _{_{\mathrm{SD}}}=9.46\pm 0.44$~mb at
1800~GeV, explicitly stating that they study proton dissociation, i.e.\ a
type~B reaction. \textbf{D0} \cite{Pal2011_D0_FPD} determined that $2\sigma
_{_{\mathrm{SD}}}=7.95\pm 0.324\pm 1.007$~mb at 1960~GeV for a type~C
reaction. While only $2\sigma _{_{\mathrm{SD}}}=2.694$~mb was directly
measured, the extrapolation into the small-$|t|$ region produced $2\sigma
_{_{\mathrm{SD}}}=7.95$~mb, while using UA4 asymptote would give $2\sigma
_{_{\mathrm{SD}}}=9.68$~mb, nearly the same as the UA4 value for $2\sigma
_{_{\mathrm{SD}}}$.

\textbf{ALICE} \cite{ALICE_Aamodt2010} initially used the $2\sigma _{_{%
\mathrm{SD}}}$ value measured by UA5 at 900~GeV, which was criticised by
Poghosyan for being inconsistently low \cite{Poghosyan2010}. Subsequently
ALICE \cite{ALICE_Abelev2013} conducted its own measurements and obtained
significantly higher results for $2\sigma _{_{\mathrm{SD}}}$ for a type~A
reaction: $11.2_{-2.1}^{+1.6}$, $12.2_{-0.2-5.3}^{+0.2+3.9}$ and $%
14.9_{+0.5-5.9}^{+0.5+3.4}$~mb at 900, 2760 and 7000~GeV,
correspondingly---results endorsed by Poghosyan \cite{Poghosyan2011} for
their consistency. Note that the upper threshold $M_{X}\leq 200$~GeV used by
ALICE in the definition of $\sigma _{_{\mathrm{SD}}}$ is different from the
conventional condition $M_{X}^{2}/s\leq 0.05$. While these definitions are
virtually identical at $\sqrt{s}=900$~GeV, ALICE's definition may
underestimate $\sigma _{_{\mathrm{SD}}}$ by $10$--$15\%$ at 2760 and
7000~GeV (model-dependent). While investigating the type~A reaction, which
is physically left--right symmetric, ALICE used advanced but asymmetric
detectors with a pseudorapidity coverage of $-3.5\leq \eta \leq 5$,
substantially skewed towards the positive-$\eta $ direction. The measured SD
events were initially left--right asymmetric, but were then corrected for
detector acceptance.

\textbf{E710} reported two major measurements specifically targeting
diffractive dissociation at 1800~GeV. These studies used the same collider
but different selection methodologies, with broader selectivity in 1990 and
stricter selectivity in 1993, resulting in different values for $2\sigma _{_{%
\mathrm{SD}}}$: $11.7\pm 2.3$~mb in 1990 \cite{E710_Amos1990} and $8.12\pm
1.7$~mb in 1993 \cite{E710_Amos1993_SD}. E710 recommended using the average
of the two measurements, $2\sigma _{_{\mathrm{SD}}}=9.4\pm 1.4$~mb, as the
best estimate for the cross-section; this value almost coincides with the
corresponding CDF measurement.

The most important feature of the E710 dataset (in the 1990 study) is that
it contains side-separated information under generally symmetric conditions.
The experiment reported $N_{L}=42904\pm 16021$ events corresponding to a
type~C reaction (antiproton dissociation) and $N_{R}=52787\pm 12582$
corresponding to a type~B reaction (proton dissociation) \cite{E710_Amos1990}%
. We first note statistical significance of the difference $\Delta
N=N_{R}-N_{L}=9883$: if a coin was thrown $N=N_{R}+N_{L}=95691$ times, then
the difference between "heads" and "tails" should be of order $\Delta N_{%
\func{std}}\sim \sqrt{N}\sim 300.$ Hence, this experiment indicates presence
of a consistent bias, which can be associated with imperfections in the
measurements or with underlying physics. The errorbars quoted by E710 are
large and point to the likelihood of a substantial systematic uncertainties (%
$\pm 12582$ and $\pm 16021$) in the measurement. These uncertainties are
factored into the displayed errorbars when $2\sigma _{\text{{\tiny B}}}$ and 
$2\sigma _{\text{{\tiny C}}}$ are shown for E710 in Figure \ref{fig2}, \
and\ when approximations for cross-sections are evaluated. We nevertheless
note that $\sigma _{\text{{\tiny C}}}/\sigma _{\text{{\tiny B}}}=\phi
^{2}\approx N_{L}/N_{R}\approx 0.812$, which corresponds to $\phi \approx
\phi _{\text{{\tiny E710}}}=0.901$.

\subsection{Approximating SD cross sections}

In line with the preceding analysis, the single-diffractive cross section, $%
2\sigma _{_{\mathrm{SD}}}$, is approximated by a Regge-motivated power law
in $s$, multiplied by the relative decoherence factor $\phi $ \textsl{as
specified in (\ref{sigSD3})}: 
\begin{equation}
2\sigma _{_{\mathrm{SD}}}^{\ast }=\sigma _{0}\,\phi ^{2k}\left( \frac{s}{%
s_{0}}\right) ^{\varepsilon },  \label{eq:fit1}
\end{equation}%
This approximation has $n_{f}=3$ fitting parameters: $\varepsilon $, $\sigma
_{0}$ and $\phi $, while $k=0,1,2,3$ denotes the effective antiproton
interaction number (Figure~\ref{fig1}), as represented by equation~(\ref%
{sigSD3}), although $k=3,$ which corresponds to reaction D, is not available
in experiments. Throughout, the squared centre-of-mass energy $s$ is
measured in $\mathrm{GeV}^{2}$; i.e. we set $s_{0}=1~\mathrm{GeV}^{2}$ so
that $(s/s_{0})^{\varepsilon }$ is dimensionless.

The approximation is fitted in the logarithmic space ($\ln \sigma _{_{%
\mathrm{SD}}}$ as a function of $\ln s$) by minimising the following
weighted deviation: 
\begin{equation}
\delta _{w}=\left( \frac{\sum_{i=1}^{n}w_{i}\ln ^{2}\left( \sigma _{i}^{\ast
}/\sigma _{i}^{\circ }\right) }{\sum_{i=1}^{n}w_{i}}\right) ^{1/2},
\label{delw}
\end{equation}%
where $\sigma _{i}^{{}}=\sigma _{i}^{\circ }\pm \delta _{\sigma i}$
represents the data, $\sigma _{i}^{\ast }$ is the corresponding
approximation, and $n$ is the number of data points. As "$\sigma $" is
conventionally used to denote cross-sections, RMS (root mean squared)
standard deviations are denoted by "$\delta $". Because the residual is
logarithmic, the statistically consistent choice is to weight by the inverse
variance in log space, 
\begin{equation}
w_{i}=\frac{1}{\delta (\ln \sigma _{i}^{\circ })^{2}}\approx \left( \frac{%
\sigma _{i}^{\circ }}{\delta _{\sigma i}}\right) ^{2},  \label{ww}
\end{equation}%
to reduce the uncertainties of the fitted parameters $\sigma _{0}$, $%
\varepsilon $ and $\phi $. Fitting is performed logarithmically so that the
dependence of $\ln \sigma _{_{\mathrm{SD}}}$ on $\ln \sigma _{0}$, $\ln \phi 
$ and $\varepsilon $ is linear.

As a secondary control, the RMS deviation measure is also evaluated using
the standard $L_{2}$ norm, defined as 
\begin{equation}
\delta _{L2}=\left( \frac{1}{n}\sum_{i=1}^{n}\left( \sigma _{i}^{\ast
}-\sigma _{i}^{\circ }\right) ^{2}\right) ^{1/2},
\end{equation}%
Note that the minimal $\delta _{w}$ does not necessarily correspond exactly
to the smallest $\delta _{L2}$ and each of these norms has its own
advantages: $\delta _{w}$ quantifies relative (multiplicative) agreement
and, with the weights above, is appropriate for determining accurate model
parameters, whereas $\delta _{L2}$ quantifies absolute deviations (in mb)
and this value tends to be more robust. While not exact, the evaluation of
parameters below shows good correspondence between the norms.

The best-fit (SDF1) parameters are 
\begin{gather}
\text{SDF1}\text{:\ \ }\ \varepsilon =0.083\pm 0.013,\ \ \sigma
_{0}=(3.66\pm 0.35)\,\mathrm{mb},\ \ \phi =0.881\pm 0.033,  \notag \\
\delta _{w}=4.1\%,\ \ \delta _{L2}=0.66\,\mathrm{mb}
\end{gather}%
where the stated uncertainty reflects both the uncertainty of the
approximation and the uncertainty of the data points. The fitted value of $%
\phi $ is in good agreement with $\phi =0.901$ previously inferred from the
direct E710 measurements. The SDF1 fit is shown by the dashed curves in
Figure \ref{fig2}.

Does the presence of parameter $\phi $ improve the approximation? If $\phi
=1 $ is enforced in equation~(\ref{eq:fit1}), the best fit becomes 
\begin{gather}
\text{SDC1}\text{:\ \ }\ 2\sigma _{_{\mathrm{SD}}}^{\ast }=\sigma
_{0}\,s^{\varepsilon },\ \ \varepsilon =0.045\pm 0.006,\ \ \sigma
_{0}=(4.79\pm 0.25)\,\mathrm{mb},  \notag \\
\delta _{w}=6.6\%,\,\ \ \delta _{L2}=1.66\,\mathrm{mb}
\end{gather}%
This SDC1 approximation is shown by the black dashed curve in Figure \ref%
{fig2}. Introducing the decoherence factor $\phi $ in SDF1 therefore
provides a substantial improvement of the quality of the approximation,
reducing the RMS deviation $\delta _{L2}$ more than 2.5 times compared to
SDC1.

Can SDF1 be improved further? Allowing $\phi $ to change with $s$ 
\begin{gather}
\text{SDF1s}\text{:\ \ \ }2\sigma _{_{\mathrm{SD}}}^{\ast }=\sigma
_{0}\,\phi _{s}^{2k}\left( \frac{s}{s_{0}}\right) ^{\varepsilon },\text{\ \
\ }\phi _{s}=\phi _{0}\left( \frac{s}{s_{0}}\right) ^{\varkappa },\ \ \
\varkappa \approx -1.5\times 10^{-4}  \notag \\
\delta _{w}=4.1\%,\qquad \delta _{L2}=0.66\,\mathrm{mb},\ 
\end{gather}%
does not appreciably affect the RMS norm or the parameters: while the $s$%
-dependent decoherence factor $\phi _{s}$ changes slightly from $\phi
_{s}\approx 0.882$ at 10GeV to $\phi _{s}\approx 0.880$ at 10000GeV. As
expected, the decoherence factor $\phi $ appears to be practically
independent of $s$.

A modest extension is to allow unequal suppression factors at successive
interaction steps, i.e.\ $\phi _{1}^{2}=\sigma _{\mathrm{B}}/\sigma _{%
\mathrm{A}}\neq \sigma _{\mathrm{C}}/\sigma _{\mathrm{B}}=\phi _{2}^{2}$.
This extension contradicts the principle reflected by equation (\ref{sigSD3}%
) but adds an additional freedom (by increasing the number of fitting
parameters from $n_{f}=3$ to $n_{f}=4$) to perform a better fit. This fit 
\begin{gather}
\text{SDF2}\text{:\ \ \ }2\sigma _{_{\mathrm{SD}}}^{\ast }=\sigma
_{0}\,s^{\varepsilon }\prod_{i=1}^{k}\phi _{i}^{2},\ \ \ \ \phi _{1}\approx
0.868,\ \ \phi _{2}\approx 0.894,  \notag \\
\delta _{w}=4.0\%,\qquad \delta _{L2}=0.77\,\mathrm{mb},\ 
\end{gather}%
produces slightly worse approximation $\delta _{L2}$. The SDF2 curves
(coloured dotted lines in Figure \ref{fig2}) are close to SDF1 --- there is
no practical reason to allow flexibility in $\phi $ beyond the prediction of
the current theory, which enforces $\phi _{1}=\phi _{2}=\phi $.

Can approximation be improved substantially without a decoherence factor? A
logarithmic control fit 
\begin{gather}
\text{SDC$\ln $}\text{:}\ \ \ \ 2\sigma _{_{\mathrm{SD}}}^{\ast }=\sigma
_{0}+\sigma _{1}\ln (s),\ \ \sigma _{0}=(4.26\pm 0.39)\,\mathrm{mb},\ \
\sigma _{1}=(0.32\pm 0.05)\,\mathrm{mb} ` \notag \\
\delta _{w}=6.6\%,\qquad \delta _{L2}=1.76\,\mathrm{mb}
\end{gather}%
is performed by minimising weighted residual in $\sigma _{_{\mathrm{SD}}}$
space, and still has a similar shape and approximation accuracy to SDC1
approximation (see Figure\ref{fig2}). The logarithmic shape was suggested by
Goulianos \cite{Goulianos1995} who obtained $\sigma _{0}=4.3\,\mathrm{mb},\
\ \sigma _{1}=0.3\,\mathrm{mb,}$ which are quite similar to the current fit.
The difference between exponential (SDC1, black dashed line in Figure \ref%
{fig2}) and logarithmic (SDCln, black dash-dotted line in Figure \ref{fig2})
approximations is not large and both can easily accommodate the decoherence
factors.

The free parameter count $n_{f}$ can be increased substantially by allowing
an energy-dependent exponent $\varepsilon =\varepsilon (s)$ but, without the
decoherence factor, this does not significantly improve the fit. For
example, if $\varepsilon (s)$ is parameterised as a fourth-order polynomial
in $\ln (s)$ and $n_{f}=5$, the fit improves to 
\begin{equation}
\text{SDC4:}\qquad 2\sigma _{_{\mathrm{SD}}}^{\ast }=\sigma
_{0}\,s^{\varepsilon (s)},\ \ \ \ \delta _{w}=5.5\%,\qquad \delta
_{L2}=1.32\,\mathrm{mb,}
\end{equation}%
but still does not match the approximation accuracy achieved by SDF1. The
SDC4 approximation, which is shown by a black dotted line in Figure \ref%
{fig2}, displays clear signs of overfitting and is impractical.

The statistical reduced chi-squared value $\chi _{w}$ that corresponds to $%
\delta _{w}$ defined by (\ref{delw}) and (\ref{ww}) is given by 
\begin{equation}
\chi _{w}^{2}=\frac{\sum_{i=1}^{n}w_{i}\ln ^{2}\left( \sigma _{i}^{\ast
}/\sigma _{i}^{\circ }\right) }{n-n_{f}}  \label{chi}
\end{equation}%
While $\chi _{w}^{2}=1.0$ is evaluated for SDC1 --- this indicates
consistency of the uncertainties estimated in experiments with approximation
SDC1--- the chi-squared value is substantially lower for SDF1: $\chi
_{w}^{2}=0.41$. From a statistical perspective, if uncertainty in estimating
standard deviations $\delta _{\sigma i}$ can be present, this indicates
overestimation of these uncertainties by a factor of $1/\chi _{w}$ where $%
\chi _{w}=0.41^{1/2}\approx 0.64,$ suggesting reduction of uncertainties of $%
n_{f}=3$ fitting parameters: $\varepsilon $, $\sigma _{0}$ and $\phi $ by
the same factor. Whether this reduction is justified or not depends, from
the statistical perspective, on the certainty of knowing exact values of $%
\delta _{\sigma i}.$ However, the reduction of the uncertainties of the
fitting parameters does not come without a cost: the tails of the
distributions become heavier (i.e. as determined by Student's $t$%
-distribution) due to variations in standard deviations.

Finally, one can neglect the uncertainties reported in experiments and set $%
w_{i}=1$ assuming $\delta _{\sigma i}=0$ in SDF1. This corresponds to the
fit 
\begin{gather}
\text{SDF1w1}\text{:\ \ }\ \varepsilon =0.080\pm 0.006,\ \ \sigma
_{0}=(3.7\pm 0.23)\,\mathrm{mb},\ \ \phi =0.888\pm 0.011,  \notag \\
\delta _{w}=7.5\%,\ \ \delta _{L2}=0.59\,\mathrm{mb}
\end{gather}%
where the uncertainties of the fitted parameters are determined solely by
the residuals of this approximation. These uncertainties are substantially
smaller than those in SDF1, for which the reported experimental
uncertainties $\delta _{\sigma i}$ dominate. The values of the fitting
parameters $\varepsilon $, $\sigma _{0}$ and $\phi $ remain nearly the same
as in SDF1. Note the similarity with the exponent $\varepsilon $ reported by
Donnachie and Landshoff \cite{DonnachieLandshoff1992} for total
cross-sections: $\sigma _{\text{tot}}=\sigma _{0}s^{\varepsilon _{0}}+\sigma
_{1}s^{-\varepsilon _{1}},\ \ \ \ \varepsilon _{0}=0.0808,\ \ \ \
\varepsilon _{1}=0.4525$

Setting $w_{i}=1$ and assuming $\delta _{\sigma i}=0$ in SDC1 approximation,
where $\phi =1$ is enforced, results in 
\begin{gather}
\text{SDC1w1}\text{:\ }\ \ \varepsilon =0.053\pm 0.012,\ \ \sigma
_{0}=(4.32\pm `0.64)\,\mathrm{mb},  \notag \\
\delta _{w}=19.1\%,\,\ \ \delta _{L2}=1.6\,\mathrm{mb}
\end{gather}%
where the value of parameter $\varepsilon $ and its uncertainty are
increased in comparison with SDC1. The detailed summary of the fits is given
in Table \ref{table1}, while MATLAB fitting and plotting routines are
provided as supplementary material.

\begin{table}[htbp]
\caption{Summary of fits for the single-diffractive cross section $2\protect%
\sigma_{SD}$. Here $\protect\delta_w$ is the weighted logarithmic RMS
deviation and $\protect\delta_{L2}$ is the RMS deviation in mb.}
\label{table1}\textsl{\ \centering
\resizebox{\textwidth}{!}{\begin{tabular}{lllllcc}
\hline
Fit & Model & $n_f$ & Main fitted parameters & $\delta_w$ (\%) & $\delta_{L2}$ (mb) & Comment \\
\hline
SDF1 &
$\sigma_0 \phi^{2k}(s/s_0)^{\varepsilon}$ &
3 &
$\varepsilon = 0.083 \pm 0.013$, $\sigma_0 = 3.66 \pm 0.35$ mb, $\phi = 0.881 \pm 0.033$ &
4.1 & 0.66 & preferred fit \\

SDF1w1 &
$\sigma_0 \phi^{2k}(s/s_0)^{\varepsilon}$ &
3 &
$\varepsilon = 0.080 \pm 0.006$, $\sigma_0 = 3.7 \pm 0.23$ mb, $\phi = 0.888 \pm 0.011$ &
7.5 & 0.59 & equal weights $w_i=1$ \\

SDC1 &
$\sigma_0 (s/s_0)^{\varepsilon}$ &
2 &
$\varepsilon = 0.045 \pm 0.006$, $\sigma_0 = 4.79 \pm 0.25$ mb &
6.6 & 1.66 & $\phi = 1$ enforced \\

SDC1w1 &
$\sigma_0 (s/s_0)^{\varepsilon}$ &
2 &
$\varepsilon = 0.053 \pm 0.012$, $\sigma_0 = 4.32 \pm 0.64$ mb &
19.1 & 1.6 & $\phi = 1$ enforced, $w_i=1$ \\

SDF1s &
$\sigma_0 \phi_s^{2k}(s/s_0)^{\varepsilon},\ \phi_s=\phi_0 (s/s_0)^\varkappa$ &
4 &
$\varepsilon = 0.084 \pm 0.014$, $\sigma_0 = 3.66 \pm 0.37$ mb, $\phi_0 = 0.882 \pm 0.15$, $\varkappa \approx -1.5 \times 10^{-4}$  &
4.1 & 0.66 & no improvement \\

SDF2 &
$\sigma_0 (s/s_0)^{\varepsilon}\prod_{i=1}^{k}\phi_i^2$ &
4 &
$\varepsilon = 0.087 \pm 0.016$, $\sigma_0 = 3.57 \pm 0.42$ mb, $\phi_1 \approx 0.868$, $\phi_2 \approx 0.894$ &
4.0 & 0.77 & slightly worse in $L2$ \\

SDCln &
$\sigma_0 + \sigma_1 \ln(s)$ &
2 &
$\sigma_0 = 4.26 \pm 0.39$ mb, $\sigma_1 = 0.32 \pm 0.05$ mb &
6.6 & 1.76 & control fit \\

SDC4 &
$\sigma_0 s^{\varepsilon(s)}$ &
5 &
$\sigma_0 = 0.16, $   $\varepsilon(s) = 1.17\xi -0.13 \xi^2 + 0.006 \xi^3 - 0.0001 \xi^4, $  $ \xi = \ln s$ &
5.5 & 1.32 & overfitting \\
\hline
\end{tabular}} }
\end{table}

\subsection{Experimental cross sections for DD collisions}

Many of the collaborations$\ $discussed in the SD section do not report the
double-diffractive cross section, $\sigma _{_{\mathrm{DD}}},$ but those who
do are of interest in this section. \textbf{ALICE}~\cite{ALICE_Abelev2013}
reports the following DD cross sections for E-type reactions (see equation~(%
\ref{rc2})): $\sigma _{\text{{\tiny E}}}=5.6\pm 2.0,\ 7.8\pm 0.2\pm 3.2,\ $%
and $9.0\pm 0.3\pm 2.6,$ for energies $\sqrt{s}$\ of\ 900, 2760 and 7000~GeV
(the second uncertainty---when quoted--is due to the luminosity). F-type
reactions are reported by \textbf{UA5}~\cite{UA5Ansorge1986} as $\sigma _{%
\text{{\tiny F}}}=(3.5\pm 2.2)\,\mathrm{mb}$ and $(4.0\pm 2.5)\,\mathrm{mb}$
at $200$ and $900~\mathrm{GeV}$, respectively, and by \textbf{CDF}~\cite%
{CDF_DD_2001} as $\sigma _{\text{{\tiny F}}}=(3.42\pm 0.01\pm 1.09)\,\mathrm{%
mb}$ and $(4.43\pm 0.02\pm 1.18)\,\mathrm{mb}$ at $630$ and $1800~\mathrm{GeV%
}$, respectively.

\begin{figure}[h]
\begin{center}
\includegraphics[width=14cm,page=1,trim=1.5cm 4cm 1.5cm 8cm, clip ]{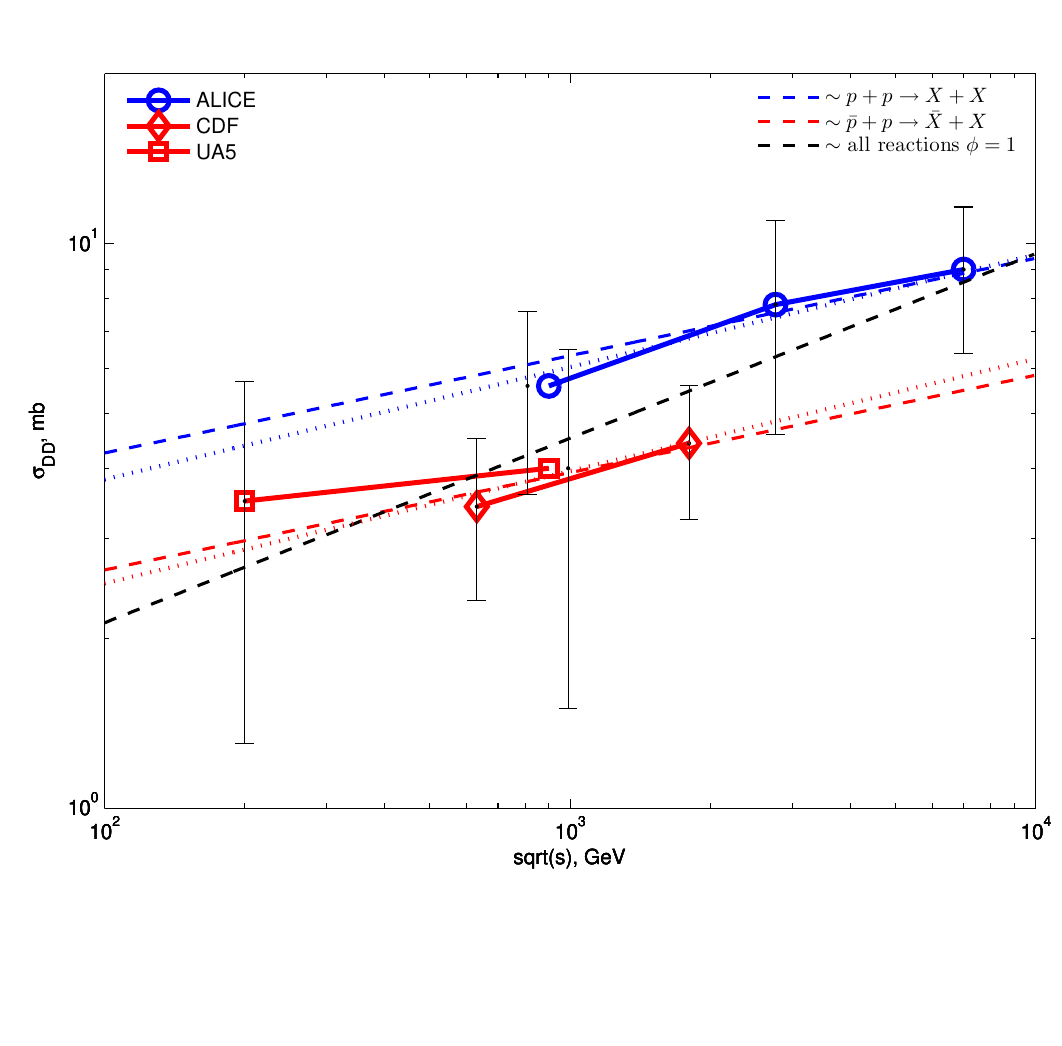}
\caption{DD cross-section $\sigma _{\mathrm{DD}}$ vs $\sqrt{(s)}$. Experimental data are from  UA5,
CDF and ALICE collaborations. Approximations: \  -- \ -- \ -- DDF1 (colour)
or DDC1 (black),\ \  $\cdot $ $\cdot $ $\cdot $ $\cdot $ $\cdot $ $\cdot \ $
DDF2}
\label{fig3}
\end{center}
\end{figure}

Figure \ref{fig3} compares these DD data with several approximations. A
Regge-motivated power of $s$ is multiplied by the relative decoherence
factor $\phi $ specified by equation~(\ref{sigDD2}), 
\begin{equation}
\sigma _{_{\mathrm{DD}}}^{\ast }=\sigma _{2}\,\phi ^{4k}\,\left( \frac{s}{%
s_{0}}\right) ^{\varepsilon _{2}},  \label{sig_DD}
\end{equation}%
where the fitting parameters are $\varepsilon _{2}$, $\sigma _{2}$ and $\phi 
$, while $k=0$ for E-type and $k=1$ for F-type reactions. In this section, $%
s $ is measured in $\mathrm{GeV}^{2}$.

We first fix $\phi =0.881$, as determined in the SD fit (SDF1), and obtain 
\begin{gather}
\mathrm{DDF1}\mathrm{:}\ \ \varepsilon _{2}=0.082\pm 0.08,\ \ \sigma
_{2}=e^{0.74\pm 1.1}\,\mathrm{mb,}  \notag \\
\delta _{w}=6.4\%,\ \ \delta _{L2}=0.47\,\mathrm{mb}.
\end{gather}%
Approximation DDF1 is shown in Figure~3 by the blue ($k=0$) and red ($k=1$)
curves. Note the broad similarity between SD's $\varepsilon $, DD's $%
\varepsilon _{2}$ and \ $\varepsilon =0.0808$\ reported for total
cross-sections by Donnachie and Landshoff \cite{DonnachieLandshoff1992}.

Allowing $\phi $ to vary in (\ref{sig_DD}) yields 
\begin{gather}
\mathrm{DDF2}\mathrm{:}\ \varepsilon _{2}=0.1\pm 0.1,\ \sigma
_{2}=e^{0.42\pm 2}\,\mathrm{mb},\ \ \phi =\phi _{_{\mathrm{DD}}}=0.899\pm
0.1,  \notag \\
\ \delta _{w}=5.6\%,\ \delta _{L2}=0.32\,\mathrm{mb}.
\end{gather}%
This slightly reduces the deviation norm from $6.4\%$ to $5.6\%$, although
the fitted $\phi $ remains close to the value inferred from SD cross
sections. Both DDF1 and DDF2 have similar uncertainties.

For comparison, a conventional power-law fit without the decoherence factor, 
\begin{gather}
\mathrm{DDC1}\mathrm{:}\ \sigma _{_{\mathrm{DD}}}^{\ast }=\sigma
_{2}\,s^{\varepsilon _{2}},\ \ \varepsilon _{2}=0.16\pm 0.1,\ \ \sigma
_{2}=e^{-0.75\pm 1.5}\mathrm{mb},  \notag \\
\delta _{w}=18\%,\ \ \delta _{L2}=0.93\,\mathrm{mb}\ 
\end{gather}%
\qquad has a noticeably worse approximation norm (black dashed curve in
Figure \ref{fig3}).

Overall, the inclusion of the decoherence factor improves the description of
the available DD data and yields the parameter $\varepsilon _{2}$ and, most
importantly, the relative decoherence factor $\phi $ consistent with those
obtained from SD cross sections. The detailed summary of the DD fits is
given in Table \ref{table2}.

\begin{table}[htbp]
\caption{Summary of fits for the double-diffractive cross section $\protect%
\sigma_{DD}$. Here $\protect\delta_w$ is the weighted logarithmic RMS
deviation and $\protect\delta_{L2}$ is the RMS deviation in mb.}
\label{table2}\textsl{\ \centering
\resizebox{\textwidth}{!}{\begin{tabular}{lllllcc}
\hline
Fit & Model & $n_f$ & Main fitted parameters & $\delta_w$ (\%) & $\delta_{L2}$ (mb) & Comment \\
\hline
DDF1 &
$\sigma_2 \phi^{4k}(s/s_0)^{\varepsilon_2}$ &
2 &
$\varepsilon_2 = 0.082 \pm 0.08$, $\sigma_2 = e^{0.74 \pm 1.1}\,\mathrm{mb}$, ($\phi=0.881$ fixed) &
6.4 &
0.47 &
$\phi$ fixed from SD fit \\

DDF2 &
$\sigma_2 \phi^{4k}(s/s_0)^{\varepsilon_2}$ &
3 &
$\varepsilon_2 = 0.10 \pm 0.10$, $\sigma_2 = e^{0.42 \pm 2}\,\mathrm{mb}$, $\phi=0.899 \pm 0.10$ &
5.6 &
0.32 &
slightly better fit \\

DDC1 &
$\sigma_2 (s/s_0)^{\varepsilon_2}$ &
2 &
$\varepsilon_2 = 0.16 \pm 0.10$, $\sigma_2 = e^{-0.75 \pm 1.5}\,\mathrm{mb}$ &
18.0 &
0.93 &
$\phi=1$ enforced \\
\hline
\end{tabular}} }
\end{table}

\bigskip

\section{Discussion}

This work yields a theoretical prediction by extending perturbative and
discrete symmetry analysis from the von Neumann to the Lindblad framework.
These results allow us to improve the approximation of single-diffraction
(SD) cross-sections and to fit measurements from the ISR, UA4, UA5, CDF, D0,
and ALICE experiments using three fitting parameters, achieving a
root-mean-square deviation of only $0.66\,\mathrm{mb}$ ($4.1\%$). This is
well below $\delta \sigma _{\mathrm{exp}}\approx 1.8\,\mathrm{mb}$, which is
the RMS average of the quoted experimental uncertainties for $\sigma _{_{%
\mathrm{SD}}}$ across the datasets considered and $1.66$ $\mathrm{mb}$ which
is RMS deviation of the conventional approximation ignoring the decoherence
factors. This quality of approximation cannot be achieved when $\phi =1$
even if a substantially large number of fitting parameters is allowed
(approximation SDC4). The main aim, however, is not merely to approximate $%
\sigma _{_{\mathrm{SD}}}$, but to estimate the relative decoherence factor $%
\phi $. This factor takes the value $\phi =1$ in CP-invariant dephasing,
whereas CPT-invariant dephasing predicts $\phi <1$. The SD, DD, and direct
event counts reported by E710 yield consistent values around $\phi \sim 0.89$%
, expected in CPT-invariant dephasing; these different, practically
independent evaluations ($\phi _{\text{{\tiny E710}}},$ $\phi _{1},$ $\phi
_{2}$ and $\phi _{_{\mathrm{DD}}}$) give close values with the relative
scattering of $\sim $1.5\%$.$It is worthwhile to note that approximations
involving the decoherence factor $\phi $ result in consistent exponent $%
\varepsilon \approx 0.08$ in $\sigma \sim s^{\varepsilon }$. Besides
CPT-invariant dephasing, which makes precisely this prediction, could there
be alternative explanations for the experimental observations that $\sigma _{%
\text{{\tiny B}}}/\sigma _{\text{{\tiny A}}}$ and $\sigma _{\text{{\tiny C}}%
}/\sigma _{\text{{\tiny B}}}$ are equal ($=\phi ^{2}\sim 0.8$) and both fall
below unity?

\bigskip

Here it is important to distinguish two comparisons with different degrees
of theoretical constraint. (i) In comparing $pp$ with $p\bar{p}$ collisions,
standard Regge phenomenology does not require exact equality at finite
energies: subleading crossing-odd exchanges and other non-asymptotic effects
can generate differences, although these are expected to diminish with
increasing $\sqrt{s}$ as vacuum exchange becomes dominant. Thus, a finite-$%
\sqrt{s}$ difference between $pp$ and $p\bar{p}$ cross-sections would be
unexpected in the leading asymptotic picture, but is not excluded in
principle. (ii) By contrast, within a single $p\bar{p}$ experiment one may
compare diffractive dissociation on the proton side with diffractive
dissociation on the antiproton side (i.e. proton dissociation with an intact
antiproton versus antiproton dissociation with an intact proton). For strong
interactions, CP and T invariance of QCD, together with symmetry under
exchanging the beam directions, implies that at fixed kinematics these two
one-sided processes should have identical statistics.

Within Regge-based approaches, where the leading crossing-even contribution
associated with Pomeron exchange dominates, one therefore expects $\sigma _{%
\text{{\tiny B}}}/\sigma _{\text{{\tiny A}}}\rightarrow 1$ at sufficiently
high energies as the secondary, crossing-odd Reggeon terms become suppressed 
\cite{DonnachieLandshoff1992,PDGSoftQCD2021}. Departures from unity are not
forbidden: for example, differences in differential elastic cross-sections
have been interpreted in terms of Odderon (leading crossing-odd Reggeon)
effects at high energies \cite{D0TOTEMOdderon2021}. In addition, at higher
energies secondary particle production can partially fill rapidity gaps and
bias diffractive identification; in principle, the size of such biases could
differ between $pp$ and $p\bar{p}$ datasets. While the author is not aware
of a consistent mechanism that robustly predicts $\sigma _{\text{{\tiny B}}%
}<\sigma _{\text{{\tiny A}}}$, such mechanisms may exist or may emerge in
future theoretical developments.

The situation is qualitatively different for the inequality $\sigma _{\text{%
{\tiny C}}}<\sigma _{\text{{\tiny B}}}$, which constitutes an apparent CP
violation (according to the event classes defined by (\ref{rc1})). Since the
strong interaction is assumed to be T-symmetric at the fundamental level 
\cite{PDGConservation2020}, such CP violation (reaching the magnitude of
10-20\%) would imply CPT violation and therefore cannot be accommodated
within standard QCD without revisiting underlying fundamental assumptions.
By contrast, random and environmental interferences provide a natural route:
non-unitary, entropy-changing processes are not T-symmetric and are
ultimately tied to the thermodynamic arrow of time. CPT-invariant dephasing
predicts $\sigma _{\text{{\tiny C}}}<\sigma _{\text{{\tiny B}}}$ and thus
permits apparent CP violations while maintaining strict CPT symmetry.
Identifying an alternative explanation within the usual local, unitary
framework would be challenging. Note the generality of the issue:
environmental interferences can readily produce apparent CPT or CP
violations even when these invariances are actually preserved within the
system \cite{K-PhysA}.

\bigskip

Finally, we note that the value of $\phi =0.881$ inferred from SD exeriments
(SDF1) could, in principle, be a coincidence arising from an unfortunate
combination of errors. This is possible. Under the quoted experimental
uncertainties, the $p$-value for the null hypothesis $\phi \geq 1$ is below $%
0.05\%$ (i.e.\ $\approx 4\times 10^{-4}$), with the Gaussian equivalent of $%
3.35$ standard deviations. If these uncertainties are rescaled using
standard procedures based on $\chi _{w}^{2}$ (\ref{chi}), the $p$-value
falls below $0.005\%$ (i.e.\ $\approx 3\times 10^{-5}$), corresponding to $%
\approx 4.0$ standard deviations. If the reported experimental errors are
neglected, the residual uncertainty ($p$-value) of approximation SDF1w1
decreases further to \ $\approx 1.2\times 10^{-8}$, corresponding to $%
\approx 5.5$ standard deviations. It is therefore clear that the reported
experimental errors dominate the overall uncertainty. \textsl{The reported }$%
p$\textsl{-values are evaluated in }$\ln (\phi )$\textsl{\ space as these
values tend to be more conservative than those corresponding to }$\phi
=0.881\pm 0.033$\textsl{\ for SDF1 and }$\phi =0.888\pm 0.011$\textsl{\ for
SDF1w1. }

The $p$-value for the null hypothesis that $\sigma _{_{\mathrm{SD}}}$ is
independent of $\phi $ is treated as two-sided and, therefore, twice as
large, i.e.\ $\lesssim 0.1\%$, $\lesssim 0.01\%$, and $\lesssim 0.000003\%$,
respectively. Outcomes with such small probabilities are conceptually
possible but, from a statistical perspective, unlikely: $p$-values in the $%
10^{-3}$--$10^{-8}$ range constitute strong evidence. When expressed as a
Gaussian-equivalent significance, these correspond to a multi-sigma effect
within the range of 3.3-5.5 standard deviations. This level of confidence is
driven by collective properties of the combined dataset, despite the
substantial uncertainty associated with each individual data point. Although
there is no evidence of major experimental mistakes or omissions, one cannot
exclude systematic measurement biases that persist across multiple
experiments and could, in principle, influence the overall outcome. The
value of repeating key measurements therefore becomes evident.
Unfortunately, the relevant experiments were performed in the past and are
no longer active. Nevertheless, at this stage---and subject to the
limitations noted above---the available experimental data favour
CPT-invariant dephasing with a substantial degree of confidence. Although
the present work is restricted to consideration of decoherence and does not
consider thermalisation and thermodynamic effects as such, we still note
that CPT-invariant dephasing broadly points towards CPT-invariant
thermodynamics \cite%
{KlimenkoMaas2012,KlimenkoMaas2014,Klimenko2017KineticsCPT,Klimenko2021b,Etesi2022,Klimenko2026Antisystems}%
.

\bigskip

A first priority would be a dedicated comparison of reactions B and C in $p%
\bar{p}$ collisions, with careful validation of detector symmetry, including
reversal of proton and antiproton directions, together with a corresponding
SD measurement in $pp$ collisions. Moderate centre-of-mass energies would be
sufficient, provided high-rapidity coverage is available. A more remote but
particularly important goal would be a comparison of $pp$ and $\bar{p}\bar{p}
$ collisions; the latter is conceptually feasible, but may be rather
challenging experimentally.

\section{Conclusion}

\textsl{This work extends the general odd-symmetric formulation---based on
stochastic realism and dual temporal conditions---to dissipative quantum
mechanics, thereby yielding a Lindblad-based time-symmetric framework that
covers contributions from both intrinsic and environmental dephasing.
Applying this framework to diffractive dissociations, together with
fundamental symmetry considerations, indicates that the corresponding
observables acquire a multiplicative relative decoherence factor }$\phi $%
\textsl{.}

Across the combined SD (single diffraction) and DD (double diffraction)
datasets, the preferred value $\phi \simeq 0.89$ yields a markedly improved
description of the measured cross-sections compared with the unitary
baseline $\phi =1$, while remaining consistent across different SD and DD
reaction classes (A, B, C, E and F) and with an independent estimate
inferred from the side-separated E710 event counts. The finding points, with
substantial statistical certainty, to the fundamental presence of intrinsic
dephasing in diffractive dissociations and to the CPT invariance of that
dephasing.

\appendix

\section{The dephasing Lindblad equation preserving energy}

\textsl{This Appendix proves that, for pure dephasing, the Lindblad
operators must commute with the Hamiltonian. }Consider the pure-dephasing
Lindblad equation \cite{Lindblad1976} 
\begin{eqnarray}
\frac{d\rho }{dt} &=&\mathcal{L}(\rho )  \label{L1} \\
&=&-\frac{i}{\hbar }[H,\rho ]+\frac{1}{\hbar ^{2}}\sum_{j}\gamma
_{j}\,\left( L_{j}\rho L_{j}^{\dag }-\frac{1}{2}\left\{ L_{j}^{\dag
}L_{j},\rho \right\} \right)  \notag \\
&=&-\frac{i}{\hbar }[H,\rho ]-\frac{1}{2\hbar ^{2}}\sum_{j}\gamma
_{j}\,[L_{j},[L_{j},\rho ]]  \notag
\end{eqnarray}%
with Hermitian $H=H^{\dag }$ and $L_{j}=L_{j}^{\dag }$, and positive rates $%
\gamma _{j}>0$ for all $j$.

Define the adjoint superoperator $\mathcal{L}^{\dagger }$ by $\func{Tr}%
\!\left( X_{1}\,\mathcal{L}(X_{2})\right) =\func{Tr}\!\left( \mathcal{L}%
^{\dagger }(X_{1})\,X_{2}\right)$ for all $X_{1}$ and $X_{2}$. For Hermitian 
$H$ and $L_{j}$, this adjoint is 
\begin{equation}
\mathcal{L}^{\dagger }(X)=\frac{i}{\hbar }[H,X]-\frac{1}{2\hbar ^{2}}%
\sum_{j}\gamma _{j}\,[L_{j},[L_{j},X]]  \label{Ladj}
\end{equation}
for any $X$.

The expectation value of the energy $E_{\rho }\overset{\text{{\tiny def}}}{=}%
\func{Tr}(H\rho )$ evolves as 
\begin{equation}
\frac{dE_{\rho }}{dt}=\frac{d}{dt}\func{Tr}(H\rho )=\func{Tr}\left( H\frac{%
d\rho }{dt}\right) =\func{Tr}\left( H\mathcal{L}(\rho )\right) =\func{Tr}%
\left( \mathcal{L}^{\dagger }(H)\rho \right)  \label{dHdt}
\end{equation}%
In the case of pure dephasing the energy must be preserved by the evolution
so that $dE_{\rho }/dt=\func{Tr}\left( \mathcal{L}^{\dagger }(H)\rho \right)
=0$. The condition $\func{Tr}\left( \mathcal{L}^{\dagger }(H)\rho \right) =0$
is valid for all density matrices $\rho $ if and only if \cite%
{Albert2014ConservedLindblad} 
\begin{equation}
\mathcal{L}^{\dagger }(H)=0.  \label{LdagH0}
\end{equation}%
Using $[H,H]=0$ in (\ref{Ladj}), condition (\ref{LdagH0}) reduces to 
\begin{equation}
\sum_{j}\gamma _{j}\,\left[ L_{j},\left[ L_{j},H\right] \right] =0.
\label{double_comm_sum}
\end{equation}%
To show that (\ref{double_comm_sum}) implies $[H,L_{j}]=0$ for every $j$,
take the Hilbert--Schmidt inner product of (\ref{double_comm_sum}) with $H$: 
\begin{equation}
0=\sum_{j}\gamma _{j}\,\func{Tr}\left( H\,\left[ L_{j},\left[ L_{j},H\right] %
\right] \right) .  \label{sum_Q}
\end{equation}%
For Hermitian $H$ and $L_{j}$, one has the identity 
\begin{equation}
\func{Tr}\left( H\,\left[ L_{j},\left[ L_{j},H\right] \right] \right) =\func{%
Tr}\left( \left[ L_{j},H\right] ^{\dagger }\left[ L_{j},H\right] \right)
=\Vert \left[ L_{j},H\right] \Vert ^{2}\geq 0.  \label{Q_identity}
\end{equation}%
The fact that the product $\left[ L_{j},H\right] ^{\dagger }\left[ L_{j},H%
\right] $ is positive semidefinite was used by Lindblad in his original
analysis \cite{Lindblad1976}.\ Applying (\ref{Q_identity}) to (\ref{sum_Q}),
we conclude that, since each $\gamma _{j}>0$ and each norm square is
nonnegative, every term must vanish: 
\begin{equation}
\lbrack L_{j},H]=0\qquad \text{for all }j.  \label{HL}
\end{equation}%
This proves the following statement.

\begin{proposition}
In the dephasing Lindblad equation (\ref{L1}) with Hermitian $H$ and $L_{j}$
and $\gamma _{j}>0$, conservation of the energy expectation holds for any
density matrix $\rho$ if and only if $[H,L_{j}]=0$ for all $j$.
\end{proposition}

If the coefficients $\gamma _{j}$ are allowed to change sign, as in the
bi-directional model (i.e. beyond the standard positivity setting), then (%
\ref{double_comm_sum}) must be enforced separately for the subsets with $%
\gamma _{j}>0$ and with $\gamma _{j}<0$.

\bigskip

\section{Temporal invariance of Lindblad operators}

\textsl{The block-diagonal structure of the Lindblad operators in the case
of pure dephasing is examined, noting covariant properties of these
operators. }We consider the eigenbasis $\left\vert k\right\rangle $, $%
k=1,2,\ldots ,n$, of the system Hamiltonian $H$, i.e. $H\left\vert
k\right\rangle =E_{k}\left\vert k\right\rangle $. If the problem involves
perturbations $H=H_{0}+H_{1}$, then the eigenstates of the principal
Hamiltonian $H_{0}$ are considered. The\ Hamiltonian is assumed to be
T-invariant so that 
\begin{equation}
\Theta _{\text{T}}H\Theta _{\text{T}}^{-1}=\mathrm{U}_{\text{T}}\mathrm{K}H%
\mathrm{KU}_{\text{T}}^{\dag }=\mathrm{U}_{\text{T}}H^{\ast }\mathrm{U}_{%
\text{T}}^{\dag }=H
\end{equation}%
hence $\mathrm{U}_{\text{T}}H^{\ast }=H\mathrm{U}_{\text{T}}$ and 
\begin{equation}
0=\left\langle i\right\vert \mathrm{U}_{\text{T}}H^{\ast }\left\vert
k\right\rangle -\left\langle i\right\vert H\mathrm{U}_{\text{T}}\left\vert
k\right\rangle =\left\langle i\right\vert \mathrm{U}_{\text{T}}\left\vert
k\right\rangle \left( E_{k}-E_{i}\right)
\end{equation}%
that is, $\left\langle i\right\vert \mathrm{U}_{\text{T}}\left\vert
k\right\rangle =0$ as long as the energy levels are not degenerate.

The same conclusion holds for $\left\langle i\right\vert L_{j}\left\vert
k\right\rangle $ since (\ref{HL}) implies $L_{j}H-HL_{j}=0$ and 
\begin{equation}
0=\left\langle i\right\vert L_{j}H\left\vert k\right\rangle -\left\langle
i\right\vert HL_{j}\left\vert k\right\rangle =\left\langle i\right\vert
L_{j}\left\vert k\right\rangle \left( E_{k}-E_{i}\right)  \label{LHHL}
\end{equation}%
Therefore, the operators $\Theta _{\text{T}}$ and $L_{j}$ are block-diagonal
in the energy eigenbasis and are strictly diagonal when the energy levels
are not degenerate. Due to dephasing, the density operator also becomes
asymptotically block-diagonal: off-block coherences decay, whereas coherence
may persist inside each block invariant under the Lindbladian evolution \cite%
{Albert2014ConservedLindblad,Fagnola2019MarkovianDephasing}.

If energy levels are not degenerate $E_{k}\neq E_{i},$ then, due to (\ref%
{LHHL}), $L_{j}$ is given by its energy eigenbasis representation 
\begin{equation}
{L}_{j}=\sum_{k}a_{j}^{k}\left\vert k\right\rangle \left\langle k\right\vert
\end{equation}%
Since $H$\ is T-invariant and Hermitian while $L_{j}$ is Hermitian, then $%
(a_{j}^{k})^{\ast }=a_{j}^{k}$ and 
\begin{equation}
\Theta _{\text{T}}L_{j}\Theta _{\text{T}}^{-1}=L_{j}  \label{TLTL}
\end{equation}%
that is the Lindbladian operators ${L}_{j}$ are also T-invariant.

The impact of energy degeneracy is now examined using physical arguments. We
consider only two states with identical energies $E_{1}=E_{2}$, noting that
such a pair may form a factor space of a larger Hilbert space. Under these
conditions, it is more convenient not to explicitly eliminate the case $%
L_{0}=I,$ which is commonly eliminated in Lindblad equations, but simply
note that the unit operator $I$ does not produce dephasing.\ The covariance
of $L_{j}$ is evaluated below for three particular degenerate cases.

\subsection{Coincident degeneracy}

If $E_{1}=E_{2}$ is not a fundamental physical constraint and, in principle,
one may also have $E_{2}=E_{1}+\Delta E$, then the previous consideration
applies for any infinitesimal $\Delta E>0$. The eigenstates and other
properties of $H$ and $L_{j}$ should then be understood as the corresponding
limits $\Delta E\rightarrow 0$. This implies that $L_{j}$ are T-invariant
and compliant with (\ref{TLTL}).

\subsection{Matter--antimatter degeneracy}

Consider the space of two distinct states $\left\vert p\right\rangle $ and $%
\left\vert \bar{p}\right\rangle $ corresponding to particle $p$ and
antiparticle $\bar{p}$, assuming the CP relations 
\begin{equation}
\mathrm{U}_{\text{CP}}\left\vert p\right\rangle =\left\vert \bar{p}%
\right\rangle ,\ \ \ \mathrm{U}_{\text{CP}}\left\vert \bar{p}\right\rangle
=\left\vert p\right\rangle
\end{equation}%
while $E_{p}$ is fundamentally the same as $E_{\bar{p}}$. Physical arguments
lead to autonomous time reversal and to decoherence between particle and
antiparticle states, which motivates 
\begin{equation}
\Theta _{\text{T}}=\mathrm{IK,\ \ \ }L_{1}=\left\vert p\right\rangle
\left\langle p\right\vert ,\ \ \ L_{2}=\left\vert \bar{p}\right\rangle
\left\langle \bar{p}\right\vert
\end{equation}%
and both $L_{1}$ and $L_{2}$ satisfy (\ref{TLTL}) and, therefore, are
T-invariant.

\subsection{Spin-1/2 states}

For a Kramers pair of spin-$1/2$ states $|\uparrow \rangle $ and $%
|\downarrow \rangle $, one may take the antiunitary time-reversal operator
in the form 
\begin{equation}
\Theta _{\text{T}}=\mathrm{U}_{\text{T}}\mathrm{K},\qquad \mathrm{U}_{\text{T%
}}=i\sigma _{y},\qquad \Theta _{\text{T}}^{2}=-\mathrm{I},
\end{equation}%
If the physical requirement is no decoherence within the spin doublet, one
can choose 
\begin{equation}
L_{0}=|\uparrow \rangle \langle \uparrow |+|\downarrow \rangle \langle
\downarrow |=I
\end{equation}%
so that $[L_{0},\rho ]=0$ on this subspace and the dissipator vanishes
there. If decoherence is allowed in all possible directions, then 
\begin{equation}
L_{0}=I,\ \ L_{1}=\sigma _{x},\ \ L_{2}=\sigma _{y},\ \ L_{3}=\sigma _{z}
\end{equation}%
where $\sigma _{x}$, $\sigma _{y}$ and $\sigma _{z}$ are the Pauli matrices.
Note that $\Theta _{\text{T}}L_{j}\Theta _{\text{T}}^{-1}=\pm L_{j}$ ($+$
for $j=0$, \ $-$ for $j=1,2,3$). This sign change, however, does not affect
the dephasing dissipator $\mathcal{D}$, which is invariant under $%
L_{j}\rightarrow -L_{j}$.

\end{document}